\documentclass[journal=jctcce,manuscript=article]{achemso}

\usepackage[version=3]{mhchem} 
\usepackage{algorithm}
\usepackage{algpseudocode}
\usepackage{physics}
\usepackage{dcolumn}
\usepackage{amsmath}
\usepackage{subfigure}
\usepackage{tikz}
\usepackage{amsfonts}
\usepackage{natbib}
\usepackage{hyperref}
\usepackage{soul}
\usepackage{bm}
\usepackage{booktabs}
\usepackage{multirow}

\renewcommand{\bra}[1]{\langle#1\rvert} 
\renewcommand{\ket}[1]{\lvert#1\rangle} 
\renewcommand{\var}[1]{\text{Var}#1} 

\newcommand{\braopket}[3]{\langle #1 | #2 | #3\rangle} 

\author{Linghua Zhu}
    \affiliation{Department of Chemistry, University of Washington, Seattle, WA 98195, U.S.A}
    \altaffiliation{Contributed equally to this work}
\author{Senwei Liang}
    \affiliation{Lawrence Berkeley National Laboratory, Berkeley, CA, 94720, U.S.A} 
    \altaffiliation{Contributed equally to this work}
\author{Chao Yang}
    \email{cyang@lbl.gov}
    \affiliation{Lawrence Berkeley National Laboratory, Berkeley, CA, 94720, U.S.A} 
\author{Xiaosong Li}
    \email{xsli@uw.edu}
    \affiliation{Department of Chemistry, University of Washington, Seattle, WA 98195, U.S.A}

\title[Running Title]{Optimizing Shot Assignment in Variational Quantum Eigensolver Measurement}

\abbreviations{IR,NMR,UV}

\begin{document}
\begin{abstract}
{Variational Quantum Eigensolvers (VQEs) show promise for tackling complex quantum chemistry challenges and realizing quantum advantages. However, in VQE, the measurement step encounters difficulties due to errors in objective function evaluation, \emph{e.g.}, the energy of a quantum state. 
While increasing the number of measurement shots can mitigate measurement errors, this approach leads to higher costs. Strategies for shot assignment have been investigated, allowing for the allocation of varying shot numbers to different Hamiltonian terms, reducing measurement variance through term-specific insights.
In this paper, we introduce a dynamic approach, the Variance-Preserved Shot Reduction (VPSR) method. This technique strives to minimize the total number of measurement shots while preserving the variance of measurements throughout the VQE process. Our numerical experiments on H$_2$ and LiH molecular ground states demonstrate the effectiveness of VPSR in achieving VQE convergence with a notably lower shot count.
}
\end{abstract}

\section{Introduction}
As quantum computers advance, we are observing continual improvements in their coherence times and gate fidelities. This encouraging trend indicates that quantum computers may soon surpass classical computers in solving some intricate and valuable real-world problems, thereby achieving quantum advantages~\cite{Peruzzo2014, Farhi2016, mcclean2016theory, cerezo2021variational, tilly2021variational,Oh2019, kochenberger2014, Lucas2014ising, Zhu2020}.
Among the numerous potential areas poised to benefit from quantum computing, the simulation of quantum many-body systems, particularly in the context of electronic structure in chemistry, stands out as a highly promising application for near-term quantum processors~\cite{whitfield2011simulation,babbush2014adiabatic,bauer2020quantum}. 

\begin{figure*}[t]
    \centering
    \includegraphics[width=6.5in]{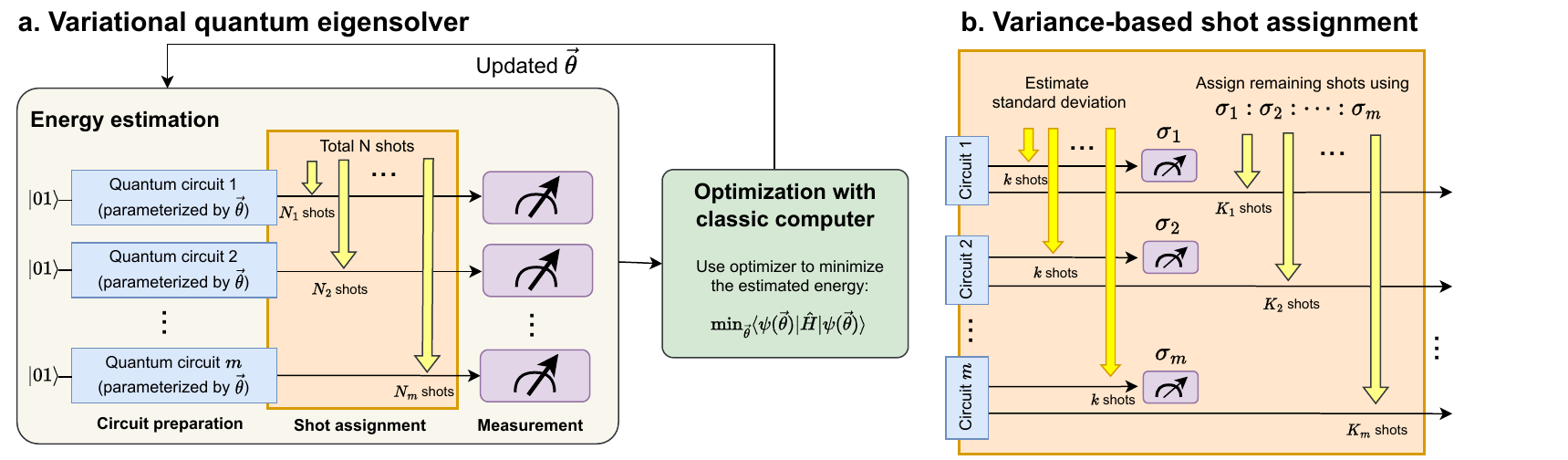}
    \caption{(a) The optimization loop in VQE. VQE involves preparing circuits parameterized by $\vec\theta$, assigning shots to measure the expectation value of the Hamiltonian, and optimizing $\vec\theta$ using classical computer. (b) On-the-fly variance estimation and variance-based assignment. To allocate the total number of $N$ shots among $m$ cliques, the standard {deviation} $(\sigma)$ of measurements is estimated by taking $k$ sampling measurement shots for each clique. The remaining measurement shots are distributed to each clique based on the estimated standard deviation. The expectation value of the energy is then computed using all measurement shots, including both the $k$ sampling shots and the optimally assigned shots.}
    \label{fig:illustration}
\end{figure*}

{Quantum many-body systems, inherently complex due to their inter-particle interactions, can be effectively studied using the variational principle.} This principle states that for any given physical system, represented by the Hamiltonian $\hat{H}$ and a wavefunction which describes a quantum state (denoted by $\ket{\psi}$), the expectation value $\braopket{\psi}{\hat{H}}{\psi}$ will always be equal to or greater than the {ground state} energy of the system.
The variational quantum eigensolver (VQE) is a well-established hybrid quantum-classical algorithm that makes use of aforementioned principle to estimate the ground state energy. VQE leverages efficient quantum evaluation of the energy function and classical optimization of circuit parameters. In such an algorithm, the wavefunction $\ket{\psi(\vec\theta)}$ is represented  by a quantum circuit parameterized by a vector $\vec\theta$. The energy expectation value associated with such a wavefunction can be obtained through a measurement process for each $\vec\theta$. The parameter $\vec\theta$ is iteratively optimized on a classical computer to minimize the energy (see Figure~\ref{fig:illustration}a)~\cite{mcclean2016theory, hadfield2019quantum,cerezo2021variational}. The advantageous features of VQE and its variants~\cite{Grimsley2019, Tang2021, tilly2021variational} have been experimentally demonstrated in many research fields~\cite{o2016scalable, mcardle2020quantum, google2020hartree}.

To extend the application of VQE to larger and more complex systems, it is crucial to address the resource-intensive nature of measurements on quantum hardware~\cite{huggins2021efficient, kiss2022quantum}. The measurement process introduces noise and perturbations, which can lead to errors in the energy objective function to be minimized. It is well known that inaccurate objective function evaluation can result in instability or slow convergence of an iterative optimization procedure carried out on classical computers~\cite{kubler2020adaptive}. To reduce the error of energy estimation, a large number of measurements (shots) are needed. These measurements are averaged to yield a more accurate energy estimation. However, taking a large number of measurements can be costly, especially when the number of terms in the Hamiltonian is large~\cite{wecker2015progress,verteletskyi2020measurement,phalak2023shot}. 
Numerous techniques have been introduced to enhance the efficiency of VQE and decrease the number of necessary measurements. Some strategies involve repartitioning Hamiltonian terms to reduce the number of {measurements}~\cite{choi2023fluid,yen2023deterministic,choi2022improving} and utilizing classical shadows for improving quantum measurements~\cite{huang2020predicting} and energy predictions~\cite{wu2023overlapped, hadfield2021adaptive, huang2021efficient, gresch2023guaranteed}. To reduce the shot count, other methods focus on refining the optimizer~\cite{luo2022koopman}, implementing tiered shot allocations during optimization~\cite{phalak2023shot}, adjusting the estimator based on past data~\cite{shlosberg2023adaptive}, and allocating variable shot counts to distinct Hamiltonian terms~\cite{arrasmith2020operator, kubler2020adaptive, gu2021adaptive}.
{Particularly, variance-minimization-based shot assignments have been successful in enhancing the accuracy and efficiency of quantum measurements~\cite{arrasmith2020operator, wecker2015progress, crawford2021efficient, zhang2023composite, yen2023deterministic, mniszewski2021reduction, williams2020understanding}. These methods typically rely on prior knowledge of Hamiltonian terms~\cite{wecker2015progress, arrasmith2020operator}, often from classical simulations, to allocate measurement shots in a manner that minimizes the overall variance of energy estimation. 
}

In this work, we introduce an on-the-fly strategy within the framework of variance minimization, namely the Variance-Preserved Shot Reduction (VPSR) approach, to optimize the quantum measurement process.
This technique is designed to work with cliques where commuting Hamiltonian terms are consolidated for simultaneous measurements. 
In addition to minimizing variance in evaluating the objective function to reduce the number of shots, this approach also considers consistent preservation of the variance during the VQE process and removes excessive measurement shots. In contrast to the pre-determined and constant shot allocation strategy, the VPSR approach also dynamically adjusts the allocation to achieve an improved shot usage efficiency.

The structure of this paper is outlined as follows: Section~\ref{sec:measure_main} provides an overview of the VQE process. This section also discusses the method of grouping various Hamiltonian terms into fewer cliques. Section~\ref{sec:strategy_overview} presents the rationale and underlying mathematical expressions of the VPSR approach for optimizing shot distribution. In Section~\ref{sec:results}, we showcase the effectiveness of  VPSR  using the VQE approach for optimizing  quantum states of the H$_2$ molecule with 2 qubits and the LiH molecule with 4 qubits. The paper concludes in Section~\ref{sec:conclusion}, where we summarize research findings and insights. The mathematical derivations of VPSR and the clique divisions of the H$_2$ and LiH molecular Hamiltonians are detailed in the appendix for further reference.

\section{Methodological Background}
\label{sec:measure_main}
\subsection{Simulating Molecular Quantum States Using the VQE Algorithm}
\label{sec:VQE_H2}

VQE is a hybrid quantum-classical algorithm designed to {search for a quantum eigenstate, often the ground state,} of a system. The VQE process is depicted in Figure~\ref{fig:illustration}a and consists of circuit preparation, shot assignment, measurement, and parameter optimization.

The initial step in computing the energy of a molecule on a quantum computer is to encode the fermionic systems onto qubits~\cite{verteletskyi2020measurement}. First, a self-consistent field method, such as the Hartree--Fock method, is often applied to obtain the single-particle orbitals as bases for many-body electronic structure methods. For a given many-body electronic structure method, such as the unitary coupled-cluster (UCC) method~\cite{bartlett1989alternative, kutzelnigg1991error, taube2006new, harsha2018difference, romero2018strategies, anand2022quantum}, represented in the second quantized form using the single-particle basis, the Jordan-Wigner~\cite{jordan1993paulische}, parity~\cite{bravyi2017tapering}, or Bravyi-Kitaev (BK)~\cite{bravyi2002fermionic} transformation can be applied to obtain the $K$-qubit Hamiltonian $\hat{H}_\text{qubit}= \sum_{i=1}^M g_{i}\hat{H}_{i}$, where $\{g_{i}\}$ represent amplitudes for $M$ terms, $\hat{H}_{i} = \bigotimes_{j=1}^{K} \sigma_{i}^{j}$ and $\sigma_{i}^{j} \in \left\{X, Y, Z, \mathbb{I} \right\}$ are the Pauli or the identity operators acting on the $j$-th qubit. The amplitudes $\{g_{i}\}$ {can be predicted from} electron integrals that depend on the molecular structures and underlying basis functions.  Additionally, certain symmetry constraints, such as conservation of particle number, fermionic parity, and spin symmetry, can be effectively used to reduce the number of qubits and terms in the Hamiltonian.~\cite{bravyi2017tapering, setia2020reducing}. 

\textbf{Circuit preparation.} Consider $U(\vec\theta)$ as a circuit parameterized by $\vec\theta$. The parameterized wavefunction, used to measure energy expectation values on a quantum processor, can be represented as:
\begin{equation}
\ket{\psi(\vec\theta)} = U(\vec\theta) \ket{\psi_{\text{ref}}},
\label{eq:wave_function}
\end{equation}
where $\ket{\psi_{\text{ref}}}$ denotes the reference state, typically selected as the Hartree--Fock solution, encompassing Hartree--Fock orbitals and {the associated electron configuration}.

\textbf{Parameter optimization.} Once we define the parameterized trial wavefunction, the expectation value of energy can be computed
\begin{equation}
  \begin{aligned}
    E(\vec\theta) &= \braopket{\psi(\vec\theta)}{\hat{H}}{\psi(\vec\theta)}
    = \braopket{\psi_{\text{ref}}}{U^{\dagger}(\vec\theta) \hat{H} U(\vec\theta)}{\psi_{\text{ref}}} \\ &
    = \sum_{i=1}^M g_{i} \braopket{\psi_{\text{ref}}}{U^{\dagger}(\vec\theta) \hat{H}_{i} U(\vec\theta)}{\psi_{\text{ref}}},
  \end{aligned}    
\label{eq:expected_energy}
\end{equation}
where the set $\{g_i\}$ denotes the coefficients corresponding to each Pauli string term $\hat{H}_{i}$ in the Hamiltonian.
The optimal value of the parameter $\vec\theta$, denoted as $\vec\theta_{\text{opt}}$, is found by solving the optimization problem 
\begin{equation}
  \begin{aligned}
\vec\theta_{\text{opt}} = \text{argmin}_{\vec\theta} E(\vec\theta).
\label{eqn:optimization}
  \end{aligned}    
\end{equation}
After evaluating the energy objective function $E(\vec\theta)$ on a quantum computer, the optimization procedure (Eq. \ref{eqn:optimization}) can be performed on a classical computer, using various optimizers, such as Adam~\cite{Kingma2014AdamAM} and the Broyden-Fletcher-Goldfarb-Shanno method~\cite{fletcher2013practical}. These optimizers are employed to iteratively update and refine the parameter $\vec\theta$ to minimize the objective function and converge towards the optimal solution.

\subsection{Consolidating Hamiltonian Terms into Cliques for Simultaneous Measurements}

A strategy to reduce the number of shots required for measuring individual Hamiltonian terms is to consolidate them into cliques, where the terms within each clique commute with each other. By simultaneously measuring the terms in a clique, the same measurement result can be used to calculate the energy contribution of each term within the clique. In contrast, terms that do not commute need to be measured separately. This approach helps reduce the overall number of shots needed for energy estimation. 

A \textit{Hamiltonian clique} refers to a subset of terms in the Hamiltonian that can be measured simultaneously. 
As illustrated in Appendix A, the Hamiltonian for H$_2$ consists of six terms, which can be consolidated into three Hamiltonian cliques because the the first three {non-identity} terms commute with each other, enabling simultaneous measurements on both qubits~\cite{crawford2021efficient, gokhale2020optimization, miller2022hardware}. While this work does not delve into the identification of these cliques, it is worth noting that determining the partition with the smallest number of cliques is a computationally challenging problem, as it is known to be NP-hard~\cite{wu2023overlapped}. Note that the proposed methods can be effective without strict requirements for the minimum number of cliques.

\section{Variance-based Assignment Strategies}
\label{sec:strategy_overview}
The measurement process described above is assumed to be deterministic and noise free. In practice, each measurement is not precise and can contain a significant amount of random noise. Therefore, each measurement of $E_i(\vec\theta)$ can be viewed as taking a sample of a random variable $\tilde{E}_i(\vec\theta)$ with a standard deviation $\sigma_i(\vec\theta)$ for the $i$th clique.

If we perform multiple quantum measurements to obtain $e_i^s(\vec\theta)$, $s=1, 2, ..., N_i$ and assume these measurements are observed samples of $\tilde{E}_i(\vec\theta)$, we can estimate $E_i(\vec\theta)$ by taking the average of $\{e_i^s(\vec\theta)\}_{s=1}^{N_i}$ to yield the empirical mean 
\begin{equation}
\bar{E}_i(\vec\theta) = \frac{1}{N_i} \sum_{s=1}^{N_i}e_i^s(\vec\theta).
\label{eq:emean}
\end{equation}
From the perspective of the Monte-Carlo sampling, we can view $\{e_i^s\}_{s=1}^{N_i}$ as realizations of independent and identically distributed random variables with same distribution associated with $\tilde{E}_i(\vec\theta)$, and view the sample mean $\bar{E}_i(\vec\theta)$ as an unbiased estimator of $E_i(\vec\theta)$. {It follows from the central limit theorem~\cite{billingsley2017probability} that 
$\sqrt{N_i}(\bar{E}_i(\vec\theta)-E_i(\vec\theta))$ 
converges to a standard normal distribution $\mathcal{N}(0,\sigma_i^2)$ as $N_i \rightarrow \infty$. This implies that the probability of $\bar{E}_i(\vec\theta)$ deviating from $E_i(\vec\theta)$ is small when $N_i$ is large.
Therefore, to reduce the effect of noise on the evaluation of the energy objective function and the convergence of the VQE, one needs to make a large of number of measurements for each clique, and estimate the value of the objective function by taking the average of these measurements.}

If the Hamiltonian terms are consolidated into $m$ cliques and $N_i$ measurements are performed for the $i$th clique, the Monte-Carlo estimate of the total energy, $\bar{E}(\vec\theta)$,
can be expressed as
\begin{align}
\bar{E}(\vec\theta):=\sum_{i=1}^m \bar{E}_i(\vec\theta)=\frac{1}{N_1}\sum_{s=1}^{N_1} e_1^s(\vec\theta)+\cdots+\frac{1}{N_m}\sum_{s=1}^{N_m} e_m^s(\vec\theta),
\label{eq:estimator}
\end{align}
For a fixed $N$, we aim to choose a partition
\begin{equation}
N = \sum_{i=1}^m N_i
\label{eq:nsum}
\end{equation}
that allows us to control {and optimize} the variance {$\var(\bar{E}(\vec\theta))$} of the estimator $\bar{E}(\vec\theta)$. 
Let $\epsilon>0$ represent the desired level of accuracy in the objective function to ensure the convergence of classical optimization. According to the Chebyshev's inequality~\cite{feller1991introduction}, the probability $\mathbb{P}\left(|\bar{E}(\vec\theta)-E(\vec\theta)|\geq \epsilon\right)$ is bounded by $\frac{\var(\bar{E}(\vec\theta))}{\epsilon^2}$. As a result, by controlling the value of $\var(\bar{E}(\vec\theta))$, we can also gain control over the probability of $\bar{E}(\vec\theta)$ deviating from $E(\vec\theta)$.

Assuming the random variables $\{\tilde{E}_i(\vec\theta)\}_{i=1}^m$ are independent, we can express the variance of  $\bar{E}(\vec\theta)$ as
\begin{align}
\var(\bar{E}(\vec\theta)) =\sum_{i=1}^m\frac{\sigma_i(\vec\theta)^2}{N_i}.
\label{eq:var}
\end{align} 
There are two main scenarios where Eq. \ref{eq:var} can aid in the optimization of shot allocation. On the one hand, when the measurement budget $N$ is limited, the distribution of $\{N_i\}$ can be optimized so that  $\sigma_i^2/N_i$, and consequently the value of the variance function, is sufficiently small. This is the goal of variance-minimization-based strategies. On the other hand, when the measurement budget $N$ is over-allocated, \emph{i.e.}, a very large number of shots, the variance of  $\bar{E}(\vec\theta)$ in Eq. \ref{eq:var} is already smaller than a target threshold. In this case, we aim to develop a new approach with the aim to minimize the number of measurements with an optimal shot allocation while keeping the variance below the target threshold (Section \ref{sec:vpsr}).

\subsection{Variance Minimization Strategy}\label{sec:vmsa}
{
To facilitate the optimization of shot allocation, a variance minimization strategy can be implemented.\cite{arrasmith2020operator, wecker2015progress, crawford2021efficient, zhang2023composite, yen2023deterministic, mniszewski2021reduction, williams2020understanding} In this paper, this type of shot assignment strategy will be broadly referred to as the Variance-Minimized Shot Assignment (VMSA) approach. This method aims to reduce the variance of the estimator (Eq. \ref{eq:estimator}), while maintaining a constant total number of shots, \emph{i.e.},
\begin{align}
\min\left\{\sum_{i=1}^m\frac{\sigma_i(\vec\theta)^2}{N_i}\right\}, \quad \sum_{i=1}^{m}N_i=N
\label{eqn:objective}
\end{align}
According to the Cauchy-Schwarz inequality, the objective (Eq. \ref{eqn:objective}) is minimized if and only if the shot allocation ratios are identical to the ratios of the standard deviations associated with the measurements of different cliques, \emph{i.e.},
\begin{align}
N_1:N_2:\cdots:N_m&=\sigma_1(\vec\theta):\sigma_2(\vec\theta):\cdots:\sigma_m(\vec\theta)
\label{eqn:ratio}
\end{align}
By allocating shots according to this ratio, one can minimize the variance of the Monte-Carlo estimate of the total energy (Eq. \ref{eq:estimator}) given a fixed shot budget, and thereby minimizing the probability that the estimation deviates from $E(\vec\theta)$. 

To obtain the ratio in Eq. \ref{eqn:ratio}, classical proxies like the amplitudes of Hamiltonian terms can be used~\cite{yen2023deterministic}.
Alternatively, standard deviation $\sigma_i(\vec\theta)$ and the variance $\sigma_i(\vec\theta)^2$ associated with each clique  can be estimated \emph{on-the-fly} using a subset of shots of the total budget ($N$). During each iteration of VQE optimization, we sample 
\begin{equation}
    e_i^s(\vec\theta)\approx \tilde{E}_i(\vec\theta), s=1,\cdots,k~(k<N/m),\notag
\end{equation}
followed by calculating the empirical variance using
\begin{equation}
\sigma_i(\vec\theta)^2\approx \frac{1}{k-1} \sum_{s=1}^k\left(e_i^s(\vec\theta)-\frac{1}{k}\sum^k_{\ell=1}e_i^\ell(\vec\theta)\right)^2. 
\end{equation}
The remaining $N-mk$ shots are distributed based on the estimated ratio (Eq. \ref{eqn:ratio}). The number of shots assigned to the $i$th clique is given by
\begin{align}
N_i=k + \frac{\sigma_i(\vec\theta)}{\sum_{j=1}^m\sigma_j(\vec\theta)}(N-mk),
\label{StD-AS}
\end{align}
where the first term ($k$) accounts for those shots used to estimate the standard deviation and the second term represents the number of additional shots assigned based on the ratio of standard deviations.
}

The number of trial shots, represented by 
$k$, is the primary hyperparameter used to estimate the standard deviation for each clique. Setting  $k$ to a low value can result in inaccurate standard deviation estimates. Conversely, a high $k$ value may constrain the remaining number of shots available for optimal allocation. As $k$ approaches $\frac{N}{m}$, the variance-minimization-based approach converges to a uniform shot assignment scheme.

\subsection{Variance-Preserved Shot Reduction Strategy}
\label{sec:vpsr}
{While minimizing variance effectively optimizes shot allocation for accelerated VQE convergence, this approach does not decrease the excessive number of measurements when the variance falls below a target threshold. To address this issue, we introduce the Variance-Preserved Shot Reduction (VPSR) method. This approach focuses on finding the optimal efficiency in shot usage while ensuring that the variance of the energy estimator remains within an acceptable limit.}

Given a threshold $\delta$ required to achieve the convergence of VQE, the objective is to solve the following optimization problem 
\begin{align}
\min_{\{N_i\}}\left\{\sum_{i=1}^{m}N_i\right\}, \quad \sum_{i=1}^m\frac{\sigma_i(\vec\theta)^2}{N_i}\leq \delta.
\label{eqn:objective2}
\end{align}
It can be shown (in Appendix B) that the minimum of the objective function in Eq. \ref{eqn:objective2} is
\begin{equation}
\Tilde{N}:=\sum_{i=1}^{m}N_i=\frac{\left(\sum_{i=1}^m\sigma_i(\vec\theta)\right)^2}{\delta }\label{eq:newN},
\end{equation}\
where $\{N_i\}$ satisfies the ratio (Eq. \ref{eqn:ratio}), and each $N_i$ takes the value of 
\begin{align}
    N_i=\Tilde{N}\frac{\sigma_i(\vec\theta)}{\left(\sum_{j=1}^m\sigma_j(\vec\theta)\right)}.
\label{eqn:vpsa2}
\end{align}

If a uniform allocation of $N$ measurement shots over $m$ cliques is known to be sufficient to ensure the convergence of VQE, the variance of the energy estimator associated with such an allocation,
\begin{equation}
\var(\bar{E}(\vec\theta)) \approx \sum_{i=1}^m \frac{\sigma_i(\vec\theta)^2}{N/m},
\label{eq:varun}
\end{equation}
   can serve as the upper bound $\delta$ in Eq. \ref{eqn:objective2}. As a result,
 the minimum of $\Tilde{N}$ becomes
 {
\begin{align}
\Tilde{N}=\frac{\left(\sum_{j=1}^m\sigma_j(\vec\theta)\right)^2}{\delta}\approx N \frac{\left(\sum_{j=1}^m\sigma_j(\vec\theta)\right)^2}{m \sum_{j=1}^m \sigma_j(\vec\theta)^2}=N\eta.
\label{eq:delta_1}
\end{align}
Eq. \ref{eq:delta_1} suggests that $\Tilde{N}\leq N$ because 
\begin{equation}
\eta=\frac{\left(\sum_{i=1}^m\sigma_i(\vec\theta)\right)^2}{m \sum_{i=1}^m \sigma_i(\vec\theta)^2}\leq 1.\label{eq:eta}
\end{equation}
This inequality can be derived from the Cauchy-Schwarz inequality,
\begin{equation}
m \sum_{i=1}^m \sigma_i(\vec\theta)^2=\sum_{i=1}^m1^2\sum_{i=1}^m \sigma_i(\vec\theta)^2\geq \left(\sum_{i=1}^m\sigma_i(\vec\theta)\right)^2.    \notag
\end{equation}
Therefore, using $\Tilde{N}$ in Eq. \eqref{eqn:vpsa2} for shot allocation will result in a reduction in the total number of quantum measurements while still achieving the same variance defined in Eq. \ref{eq:varun}. 

Similar to the way variance is estimated in the VMSA approach introduced in Section \ref{sec:vmsa}, VPSR also needs an initial set of $k$ shots to estimate the standard deviation for each clique {and $\eta$ in Eq. \ref{eq:eta}}. As a result, the value of $N$ in Eq. \ref{eq:delta_1} is replaced by $N-mk$ to account for the shots used for initial variance estimation. Therefore, in VPSR, the number of shots assigned to the $i$th clique is given by
\begin{align}
N_i=
k +  \eta\frac{\sigma_i(\vec\theta)}{\sum_{j=1}^m\sigma_j(\vec\theta)}(N-mk).
\label{StD-VPSR}
\end{align}
 Upon comparing Eq. \ref{StD-VPSR} to Eq. \ref{StD-AS} and considering $\eta\le1$, as established in Eq. \ref{eq:eta}, it becomes evident that the VPSR method consistently matches or surpasses the effectiveness of the VMSA strategy.
}

\section{Non-Variance-Based Assignment Strategies}

In this study, we also evaluate variance-based shot assignment strategies in comparison to methods that do not rely on variance.

\subsection{Uniform Assignment Strategy} 
When little is known about the noise statistics associated with the measurement, one may assume all measurement processes have the  same standard deviation, \emph{i.e.}, $\sigma_i(\vec\theta)=\sigma$ for all $i$. In this case, it may be reasonable to distribute the available measurement shots $N$ uniformly across all cliques, $N_i=\frac{N}{m}$ for all $i$. While this assumption of equal noise level may not hold in general, distributing measurement shots evenly among different clique remains effective when the total measurement budget is sufficiently large. If $\sigma_{\max} = \max_i \{\sigma_i\}$ and $\sigma_{\max}^2/(N/m)$ is sufficiently small, the total variance (Eq. \ref{eq:var}) will be small. As a result, the noise in a single  measurement will not significantly impact the convergence of optimization in VQE. However, when the measurement budget is limited, this uniform assignment is not optimal and it can lead to significant error in the evaluation of the energy objective function and inhibit the convergence of the VQE algorithm.

{
\subsection{Amplitude-Based Shot  Strategy}

Additionally, we explore the amplitude-based shot assignment (ABSA) strategy, exemplified by the ``measurement strategy with provable guarantees'' as outlined in Ref. \citenum{gresch2023guaranteed}. 
The ABSA strategy, as part of a heuristic method in conjunction with the ShadowGrouping technique, is detailed in Ref. \citenum{gresch2023guaranteed}. In this study, we have tailored the ABSA strategy to fit the experimental framework employed.

The ABSA approach  focuses on minimizing an objective function that incorporates the amplitude of each term:
\begin{align}
    \min\sum_{i=1}^M\frac{g'_i}{\sqrt{N_i}}, \quad \sum_i^M N_i=N.
    \label{eqn:amplitude}
\end{align}
where $g_i'$ is the sum of amplitudes of all terms in the $i$th clique.
Given $N$ shots, the objective is to find a measurement scheme to minimize the probability of the estimator of the energy deviating from the true energy. 

For a constant clique setting of non-overlapping groups, the minimization problem given in Eq. \eqref{eqn:amplitude} can be solved analytically. We take the H$_2$ molecule (Appendix A) as an example to illustrate how to solve such an minimization problem. For the H$_2$ molecule, the minimization problem defined in \eqref{eqn:amplitude} becomes 
\begin{align*}
    \min\left\{\frac{|g_1|+|g_2|+|g_3|}{\sqrt{N_1}}+\frac{|g_4|}{\sqrt{N_2}}+\frac{|g_5|}{\sqrt{N_3}}\right\}, \quad N_1+N_2+N_3=N.
\end{align*}
Using the H\"{o}lder's inequality, we obtain 
\begin{align*}
   \left(\sum_{i=1}^3 N_i\right)^\frac{1}{3}\left(\frac{ g_1'}{\sqrt{N_1}}+\frac{g_2'}{\sqrt{N_2}}+\frac{|g_3'|}{\sqrt{N_3}}\right)^{\frac{2}{3}}&\geq N_1^\frac{1}{3}\frac{\left(g_1'\right)^{\frac{2}{3}}}{N_1^{\frac{1}{3}}}+N_2^\frac{1}{3}\frac{\left(g_2'\right)^{\frac{2}{3}}}{N_2^{\frac{1}{3}}}+N_3^\frac{1}{3}\frac{\left(g_3'\right)^{\frac{2}{3}}}{N_3^{\frac{1}{3}}}\\&=\left(g_1'\right)^{\frac{2}{3}}+\left(g_2'\right)^{\frac{2}{3}}+\left(g_3'\right)^{\frac{2}{3}},
\end{align*}
where $g_1' = |g_1|+|g_2|+|g_3|$, $g_2' = |g_4|$ and $g_3' = |g_5|$.
The equality holds when 
\begin{equation}
    \frac{|g_1|+|g_2|+|g_3|}{N_1^{\frac{3}{2}}}=\frac{|g_4|}{N_2^\frac{3}{2}}=\frac{|g_5|}{N_3^\frac{3}{2}}.\notag
\end{equation}
Consequently, the assignment ratio  
\begin{align*}
    N_1:N_2:N_3=\left(g_1'\right)^{2/3}:\left(g_2'\right)^{2/3}:\left(g_3'\right)^{2/3} = (|g_1|+|g_2|+|g_3|)^\frac{2}{3}:|g_4|^\frac{2}{3}:|g_5|^\frac{2}{3}
\end{align*}
minimizes the objective function in Eq. \ref{eqn:amplitude}. 

This mathematical derivation shows that in the ABSA strategy, optimal shot allocation is determined by the ratios of $\left(g_i'\right)^{2/3}$.
}

\section{Numerical Results}
\label{sec:results}

{In this section, we demonstrate the efficacy of the proposed VPSR strategy through numerical results. Specifically, we optimize the ground state wavefunction of H$_2$ and  LiH molecules using the VQE method. 
The primary objective of these numerical examples is to demonstrate that (1) VPSR can maintain a desired variance level for the energy estimation, and (2) VPSR preserves the convergence to the ground state energy with a reduced number of measurement shots.}

\subsection{Experiment Setup}

All experiments were implemented with Qiskit QasmSimulator.\cite{Qiskit} We utilize the Adam~\cite{Kingma2014AdamAM} method, a stochastic gradient-based optimizer with momentum acceleration, to minimize the  objective function Eq. \ref{eq:estimator}. The learning rate used in Adam to update the parameter is initialized to $0.1$ and it follows a cosine decay, where it decreases as a factor of $0.5( \cos(\frac{\pi t}{T}) + 1)$ at iteration $t$. The gradient of the parameter is estimated by finite difference with a step size of $0.02$. 

{The variance-based assignment strategy (include VMSA and VPSR) is summarized in the pseudo code provided in Algorithm~\ref{code:radio}.}

\begin{algorithm}
    \caption{{On-the-fly} variance estimation approach}  
    \label{code:radio}
\begin{flushleft}
\textbf{Input:} Hamiltonian $H$, Ansatz $U(\vec\theta)$, initial $\vec\theta$, optimizer, shot budget $N$, and max iteration $T$. \\

\textbf{Define:} $|\psi_\text{ref}\rangle$. \\

\textbf{Prepare quantum circuit:} quantum circuit for the $j$th Hamiltonian clique
    
\textbf{Output:} $\vec\theta$. \\
\end{flushleft}
\begin{algorithmic}[1]
    \For{$t$ from 1 to $T$} 
    \State Prepare state $|\psi(\vec\theta)\rangle=U(\vec\theta)|\psi_{\text{ref}}\rangle$
    \State Estimate $\sigma_i(\vec\theta)$ with $k$ shots for $i=1,\cdots,m$
    \State Use the remaining $N-km$ shots for optimal assignments
    \State Assign $N_i$ shots to the $i$th clique using Eq. \ref{StD-AS} for VMSA or Eq. \ref{StD-VPSR} for VPSR
    \State Compute energy $\bar{E}_i(\vec\theta)$, $i=1,\cdots, m$
    \State $\bar{E}(\vec\theta)\leftarrow \sum_{i} \bar{E}_i(\vec\theta)$
    \State $\vec\theta\leftarrow {\rm optimizer}(\bar{E}(\vec\theta))$
    \EndFor
    \end{algorithmic} 
\end{algorithm}  

In the QasmSimulator, random Pauli bit flips are uniformly applied by default to introduce systematic measurement noise~\cite{van2022model}.
To better emulate the behavior of a real quantum system, we incorporated additional four types of errors: gate errors, reset errors, phase errors, and measurement errors. These errors were injected with probability $p=0.0001$~\cite{ezratty2023we}. 

\subsection{H$_2$ Molecule with Two Qubits}

{The Unitary Coupled-Cluster Single and Double (UCCSD) ansatz~\cite{barkoutsos2018quantum}, a truncated version of the UCC approach,} is an excellent choice for quantum simulation of chemical systems due to its representation as unitary transformations, enabling straightforward circuit implementation. Additionally, the parameterization of UCCSD involves only a quartic number of parameters, making it suitable for moderately sized systems without encountering significant hardware limitations. Moreover, UCCSD exhibits remarkable accuracy for non-strongly correlated systems~\cite{chen2021quantum}, further enhancing its appeal for VQE applications. 

{In this first test, numerical experiments are carried out using the UCCSD wavefunction ansatz and the minimal STO-6G basis for an H$_2$ molecule with a bond length of $1.75$ \AA. Such a wavefunction can be effectively represented by two qubits,
\cite{o2016scalable,hempel2018quantum} and  $U(\vec\theta)$ contains a single parameter $\theta$ and can be written as:
\begin{equation}
U(\theta) = \text{exp}(-i\theta X_{0}X_{1}),
\label{eq:ansatz}
\end{equation}
where $\ket{\psi_{\text{ref}}} = \ket{\text{01}}$, representing the Hartree--Fock reference of the H$_2$ system.
Each qubit represents a molecular spin-orbital: $\ket{0}$ and $\ket{1}$ denote unoccupied and occupied orbitals.} 

\begin{figure}[ht]
           \centering
           \includegraphics[width=3.in]{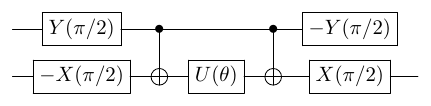}
           \caption{The quantum circuit implementation of the UCCSD wavefunction ansatz $\text{exp}(- i\theta X_{0}X_{1})$ for an H$_2$ molecule with a minimal basis.}
           \label{fig:circ_1d}
\end{figure}

Figure \ref{fig:circ_1d} shows the quantum circuit for the UCCSD state preparation. For this simple system, the initial wavefunction is prepared with $\theta=-2.0$, which yields an energy estimation far from {the ground state energy  or the Hartree--Fock reference, in order to create a sufficiently long classical optimization process that exhibits the performance of the shot optimization strategies.  We should note that, in general, one can start from $\theta = 0.0$ which corresponds to the Hartree--Fock reference.  However, for a small system such as H$_2$, the VQE optimization converges quickly from $\theta=0.0$. The rapid convergence makes it difficult to examine and compare the effects of different shot assignment strategies on problems that may require more VQE iterations to converge even from the Hartree--Fock reference.  Therefore, in this example, we choose to start from $\theta=2.0$ to illustrate the effect of shot assignment strategies on the convergence of VQE.}

The Bravyi-Kitaev transformed Hamiltonian {with spin symmetries} using two qubits for an H$_2$ molecule is
\begin{equation}
\hat{H} = g_{0} \mathbb{I} + g_{1} \mathbb{I} \otimes Z_0 + g_{2}Z_1 \otimes \mathbb{I} +g_{3}Z_{1} \otimes Z_{0} + g_{4}Y_{1}\otimes Y_{0} + g_{5}X_{1}\otimes X_{0},
\label{eq:H2_ham}
\end{equation}
where the amplitudes $\{g_i\}$ correspond to the integrals obtained from a Hartree--Fock calculation based on the bond distances and the type of atomic basis function. {$g_{1-5}$ terms are consolidated into three cliques during the quantum measurement process.} The relevant amplitudes $\{g_i\}$ in Eq. \ref{eq:H2_ham} and 
details regarding consolidating terms into cliques can be found in Appendix A.

\begin{figure}[ht]
\centering
{\includegraphics[width=.45\textwidth]{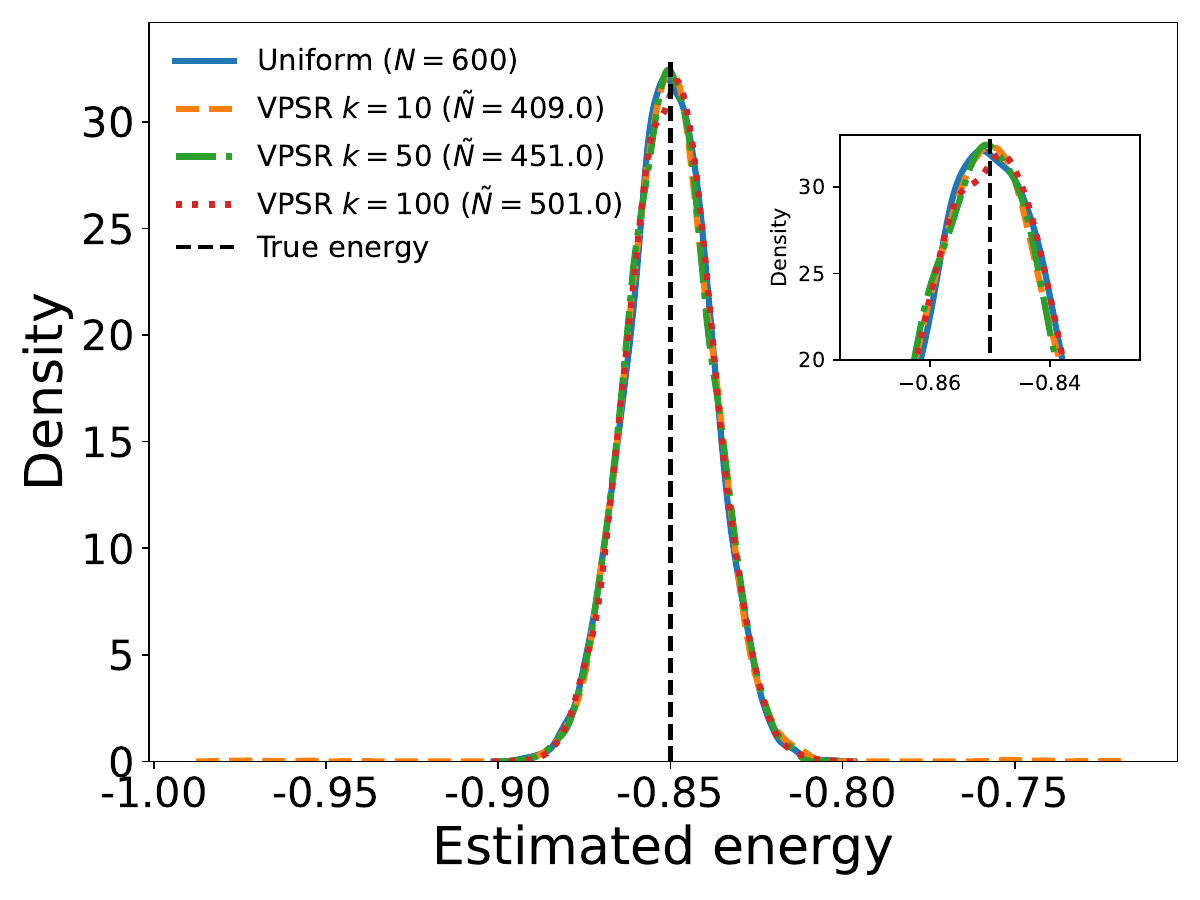}}
{\includegraphics[width=.45\textwidth]{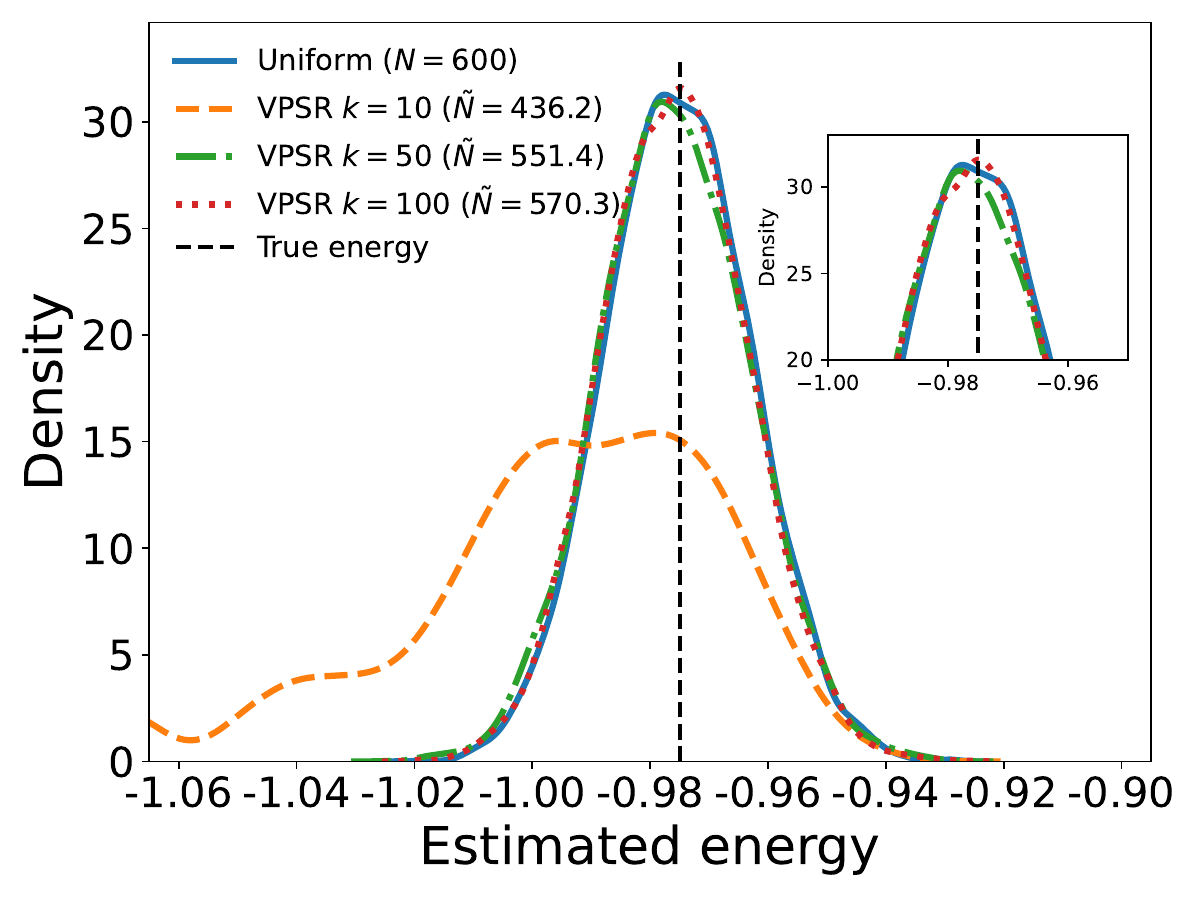}}
\caption{Distributions of the estimated energy expectation values of an H$_2$ molecule for wavefunctions {prepared} with (left) $\theta=0$  and (right) $\theta=0.91$ based on 10,000 independent experiments. }\label{fig:energyestimationH2}
\end{figure}

{As discussed in the previous section, an effective selection of the hyperparameter, \emph{i.e.}, the number of shots ($k$) for variance estimation, is important for enhancing the VQE optimization process.
To illustrate the efficacy of the VPSR approach and assist in selecting the value of $k$ during the VQE optimization process, we conducted 10,000 experiments with a total shot budget of $N=600$ for each run. In these experiments, we estimate the energy expectation value $E(\theta)$, starting from two initial states: $\theta=0$, representing the Hartree–Fock reference, and $\theta=0.91$, corresponding to the optimized UCCSD ground state wavefunction. 

Figure~\ref{fig:energyestimationH2} presents a comparison of the probability density functions of energy estimations generated using both uniform assignment and the VPSR shot allocation/reduction strategy. At $\theta=0$, the probability density distributions from all methods look similar, with the highest probability concentrated around the Hartree--Fock energy. The VPSR approach with the smallest sample size, $k=10$ (depicted by the dashed orange curve), shows the most significant reduction in measurement shots. However, it also exhibits non-zero probabilities in areas far from the center of the distribution. This deviation is more noticeable at $\theta=0.91$. On the other hand, the VPSR algorithm with both $k=50$ and $k=100$ maintains accurate control over the probability distribution, with a reduced measurement budget compared to uniform assignment. This indicates that a small sample size $k$ may result in inaccurate variance estimation. Furthermore, since variance varies with the variational parameter, different $\theta$ values should be explored to determine an appropriate $k$ value. It's important to note that in a system where the optimized variational parameters are unknown, the $k$ test can be performed using a variety of initial wavefunction guesses. Based on these results, we select $k=50$ for the VQE process.}

\begin{figure}[ht]
\centering
\includegraphics[width=.95\textwidth]{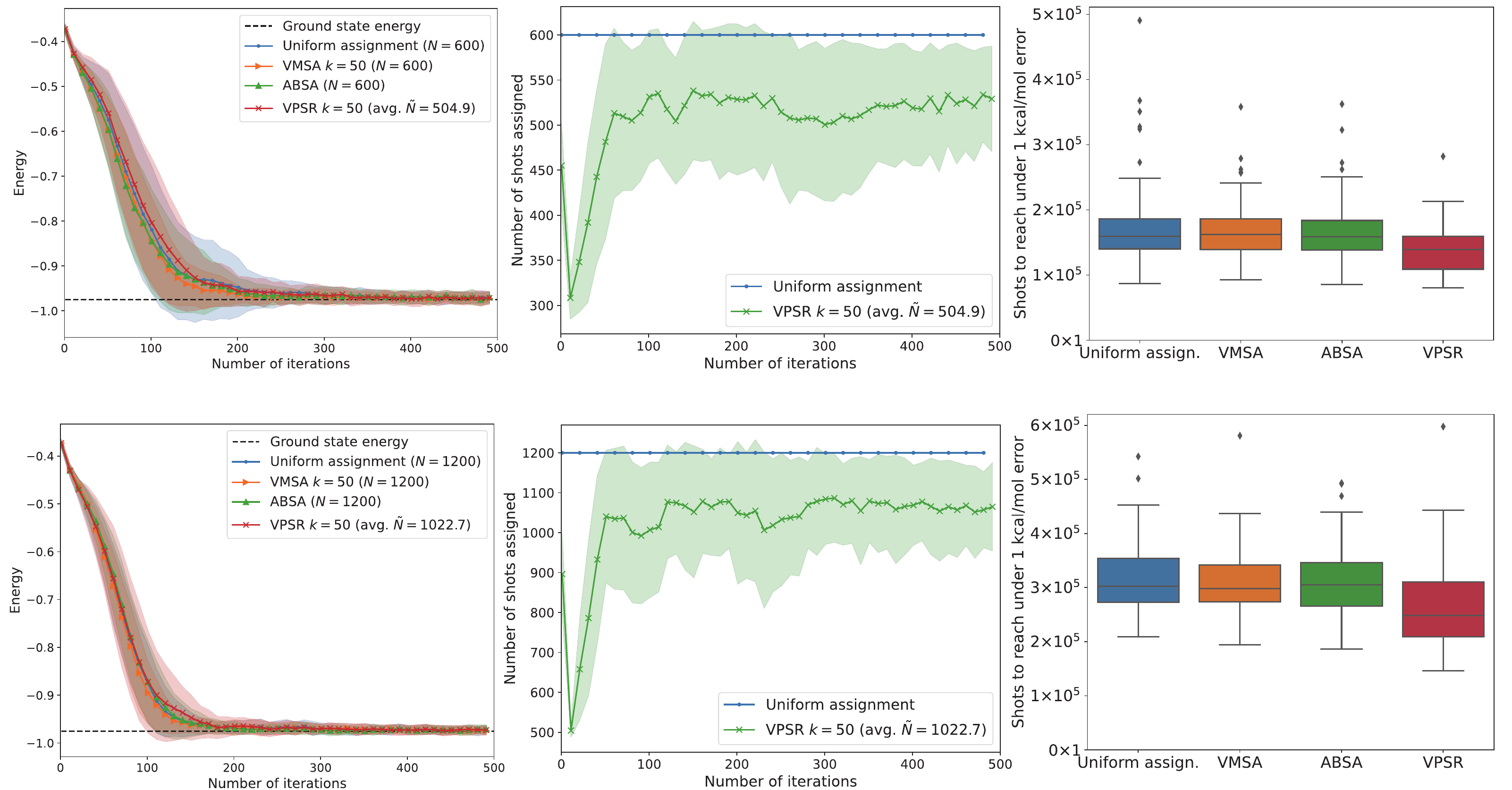}

\caption{Comparison of different shot assignment strategies for the VQE optimization of a one-dimensional $\theta$ parameter in the H$_2$ ground state wavefunction. Top Panel: $N=600$, $k=50$. Bottom Panel: $N=1200$, $k=50$. 
In the first column, the energy is plotted as a function of the number of iterations {with the standard deviation indicated by a shaded area}. The second column visualizes the distribution of shot counts across each iteration. The third column contains boxplots that summarize the number of shots needed to obtain an energy {approximation within 1.0 kcal/mol (1.6 mHartree) from the ground state energy}. These boxplots facilitate a comparison of the distributional differences between datasets. Within each boxplot, the five horizontal lines, from bottom to top, represent the minimum, 25th percentile, median, 75th percentile, and maximum values. Black dots are used to denote outlier data points. The total number of shots utilized upon achieving the desired convergence: $N=600$ -- Uniform ($1.8\times10^{5}$),
ABSA ($1.7\times10^{5}$),
VMSA ($1.8\times10^{5}$),
VPSR ($1.4\times10^{5}$); $N=1200$ --
Uniform ($3.2\times10^{5}$),
ABSA ($3.2\times10^{5}$),
VMSA  ($3.1\times10^{5}$),
VPSR  ($2.7\times10^{5}$).
}\label{fig:H2VQE}
\end{figure}

Figure~\ref{fig:H2VQE} showcases a comparison of the VPSR strategy with {the VMSA approach and non-varianc-based methods (uniform and ABSA)}, using two different shot budgets: $N=600$ and $N=1200$. In the first and second columns, each curve represents the average results from 60 independent experiments, with the standard deviation indicated by the shaded area. The third column in Figure \ref{fig:H2VQE} analyzes the total number of measurement shots necessary to attain an energy expectation value with an error under 0.5\% of the ground state energy.

Benchmark tests reveal that VPSR efficiently reduces the number of shots required while achieving a convergence profile comparable to other approaches. {For shot budgets of $N=600$ and $N=1200$, the VPSR approach achieves a reduction in the number of measurements needed for convergence (to within 1 kcal/mol from the ground state) by approximately 14-22\%, relative to the other three strategies}. The second column of Figure~\ref{fig:H2VQE} illustrates that the total shots allocated by VPSR vary throughout the VQE optimization, highlighting the adaptive nature of the shot allocation process.

\begin{figure}[ht!]
\centering
{\includegraphics[width=.45\textwidth]{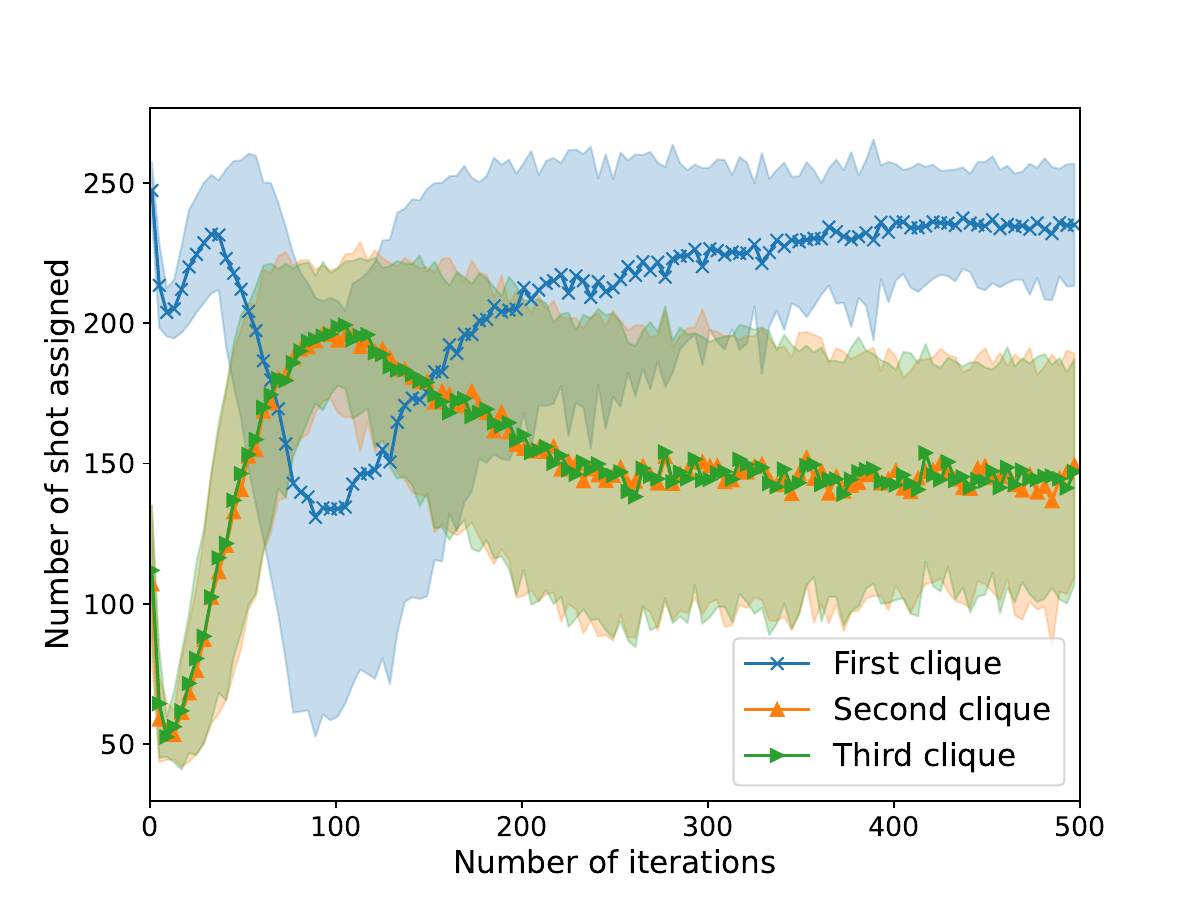}}
{\includegraphics[width=.45\textwidth]{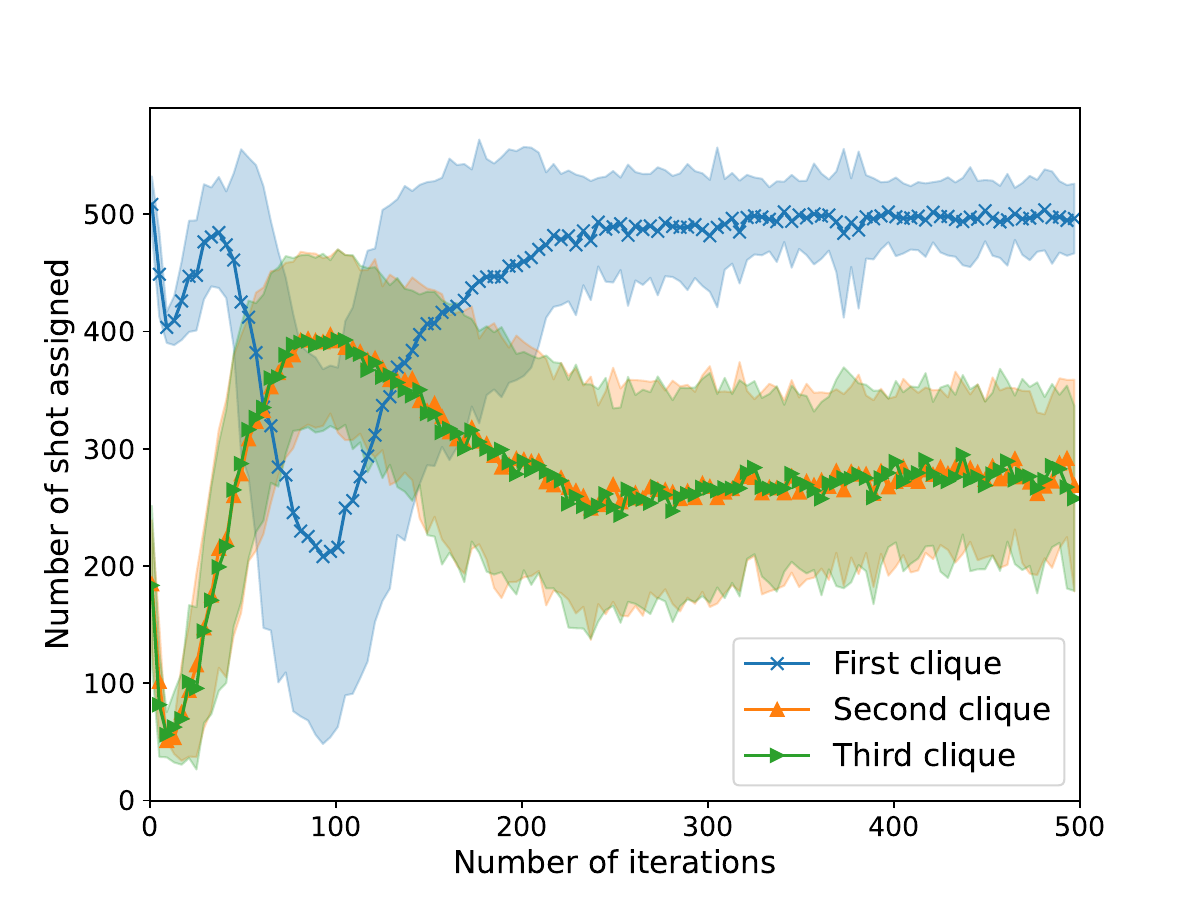}}
\caption{The number of shot assigned to each clique in VPSR for the VQE optimization of a one-dimensional $\theta$ parameter in the H$_2$ ground state wavefunction. Left: $N=600$, Right: $N=1200$. }\label{fig:H2shotassignment}
\end{figure}

Figure \ref{fig:H2shotassignment} provides insights into the adaptive nature of VPSR shot allocation throughout the VQE optimization process. Initially, more shots are assigned to the first clique than to the other two, highlighting an initial strategy of concentrating resources on optimizing the most critical clique for maximal improvement. As the optimization progresses towards convergence, this allocation trend shifts, with the first clique receiving fewer shots. Notably, in the later stages (iteration $>$150), the allocation shifts once more, with the last two cliques receiving a significantly lower number of shots than the first. This trend indicates the VPSR's capability to dynamically reallocate shots among cliques, adapting to the progresses of the VQE optimization process.

\begin{figure}[ht!]
\centering
{\includegraphics[width=.95\textwidth]{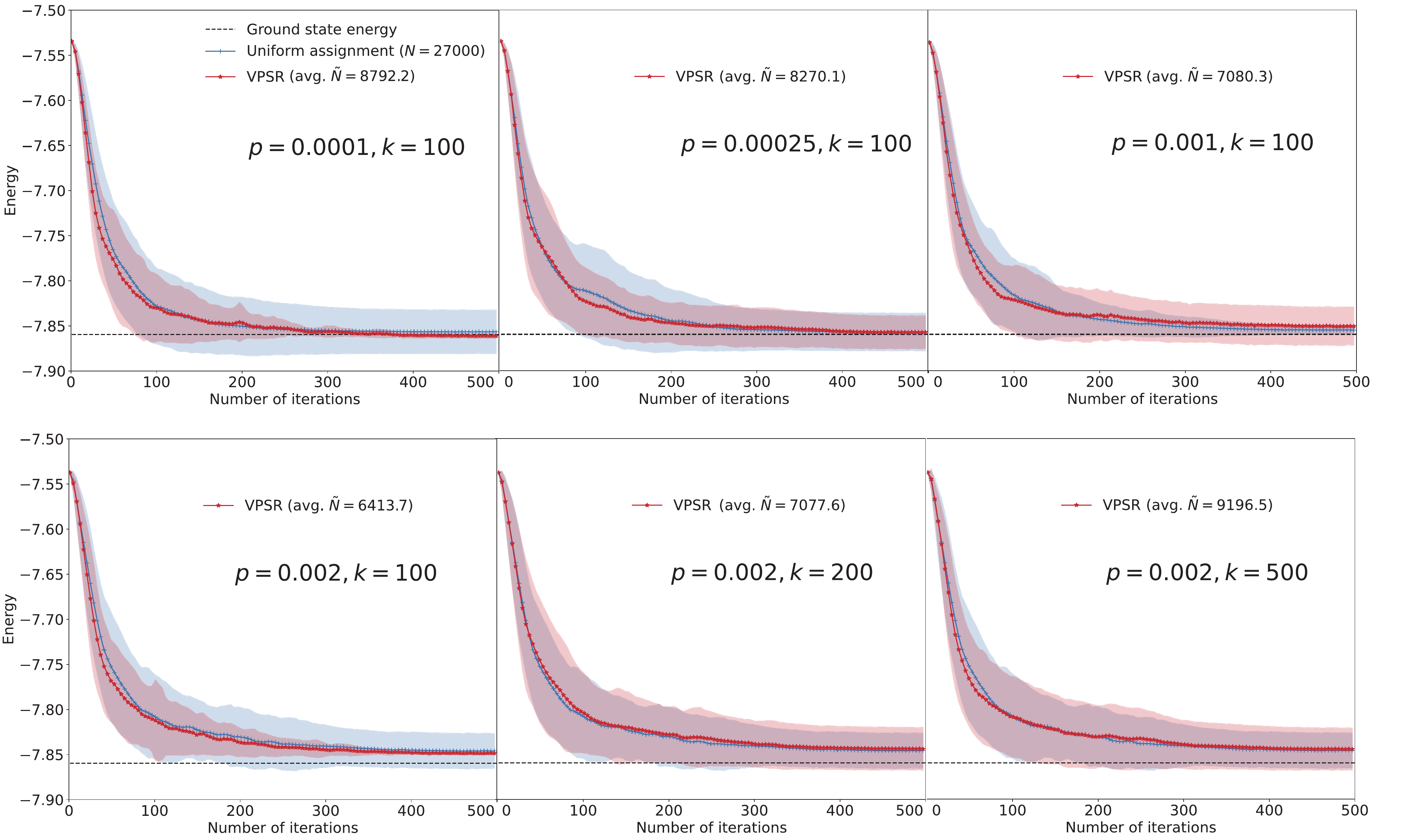}}
\caption{Comparison of the uniform and VPSR assignment strategies for the VQE optimization of an eight-dimensional $\vec\theta$ parameter in the LiH ground state wavefunction under various noise levels in measurement with different choices of $k$. To better emulate the behavior of a real quantum system, we incorporated additional four types of errors: gate errors, reset errors, phase errors, and measurement errors. These errors were injected with probability $p$. Top Left: $p=0.0001, k=100$, Top Middle: $p=0.00025, k=100$, Top Right: $p=0.001, k=100$, Bottom Left: $p=0.002, k=100$, Bottom Middle: $p=0.002, k=200$ and Bottom Right: $p=0.002, k=500$.}\label{fig:LiHnoiselevel}
\end{figure}

To test the noise resilience of the VPSR approach, we conducted experiments across multiple noise levels, the results of which are depicted in Figure \ref{fig:LiHnoiselevel}. This figure illustrates that the VPSR method retains its efficiency in shot reduction and convergence stability across these conditions. Furthermore, at the highest noise level, we explored the impact of varying the $k$ sampling size (lower panel in Figure \ref{fig:LiHnoiselevel}), finding that the convergence profile remained consistent.

\subsection{LiH Molecule with Four Qubits}

{
In a minimal basis setup, the wavefunction of lithium hydride (LiH) can be represented using 12 qubits, where each qubit corresponds to a molecular spin orbital. By implementing the Parity transformation along with symmetries and excluding non-bonding orbitals, the Hamiltonian can be effectively simplified to a form that requires only four qubits.~\cite{rattew2019domain,lolur2023reference,choy2023molecular} This simplified form is expressed as a combination of 27 Pauli strings,
\begin{align}
\begin{split}
\hat{H}_{\text {LiH }} & =g_0 \mathbb{I}+g_1 X_3 \otimes X_2 \otimes Y_1 \otimes Y_0 +g_2 X_3 \otimes Y_2 \otimes Y_1 \otimes X_0 +g_3 X_3 \otimes Z_2 \otimes X_1 \otimes \mathbb{I}\\ 
&+g_4 X_3 \otimes Z_2 \otimes X_1 \otimes Z_0
 +g_5 X_3 \otimes \mathbb{I} \otimes X_1 \otimes \mathbb{I} +g_6 Y_3 \otimes X_2 \otimes X_1 \otimes Y_0\\
 & + g_7 Y_3 \otimes Y_2 \otimes X_1 \otimes X_0 +g_8 Y_3 \otimes Z_2 \otimes Y_1 \otimes \mathbb{I} + g_9 Y_3 \otimes Z_2 \otimes Y_1 \otimes Z_0\\
& +g_{10} Y_3 \otimes \mathbb{I} \otimes Y_1 \otimes \mathbb{I} +g_{11} Z_3 \otimes \mathbb{I} \otimes \mathbb{I} \otimes \mathbb{I}+ g_{12} Z_3 \otimes X_2 \otimes Z_1 \otimes X_0 \\
& + g_{13} Z_3 \otimes Y_2 \otimes Z_1 \otimes Y_0 + g_{14} Z_3 \otimes Z_2 \otimes \mathbb{I} \otimes \mathbb{I} +g_{15} Z_3 \otimes \mathbb{I} \otimes Z_1 \otimes \mathbb{I} \\
& + g_{16} Z_3 \otimes \mathbb{I} \otimes \mathbb{I} \otimes Z_0 + g_{17} \mathbb{I} \otimes X_2 \otimes Z_1 \otimes X_0 + g_{18}\mathbb{I} \otimes X_2 \otimes \mathbb{I} \otimes X_0 \\
& + g_{19} \mathbb{I} \otimes Y_2 \otimes Z_1 \otimes Y_0 + g_{20}\mathbb{I} \otimes Y_2 \otimes \mathbb{I} \otimes Y_0 + g_{21} \mathbb{I} \otimes Z_2 \otimes \mathbb{I} \otimes \mathbb{I} + g_{22} \mathbb{I} \otimes Z_2 \otimes Z_1 \otimes \mathbb{I} \\
& + g_{23} \mathbb{I} \otimes Z_2 \otimes \mathbb{I} \otimes Z_0 + g_{24} \mathbb{I} \otimes \mathbb{I} \otimes Z_1 \otimes \mathbb{I} + g_{25}\mathbb{I} \otimes \mathbb{I} \otimes Z_1 \otimes Z_0 + g_{26} \mathbb{I} \otimes \mathbb{I} \otimes \mathbb{I} \otimes Z_0.
\end{split}
\label{eqn:lih_hamiltonian}
\end{align}
Utilizing qubit-wise commutativity, the Pauli strings in Eq. \ref{eqn:lih_hamiltonian} are grouped into 9 distinct cliques. Values of the coefficients at bond length $R = 1.45$ \AA~using the minimal STO-3G basis, along with details about the grouped cliques, are provided in Appendix C. In the experiments, for more convenient implementation on NISQ devices, we utilize a two-layer hardware-efficient ansatz.\cite{choy2023molecular} This approach leads to an eight dimensional parameter set, \emph{i.e.}, the length of $\theta$ is eight. 

\begin{figure}[ht]
\centering
{\includegraphics[width=.45\textwidth]{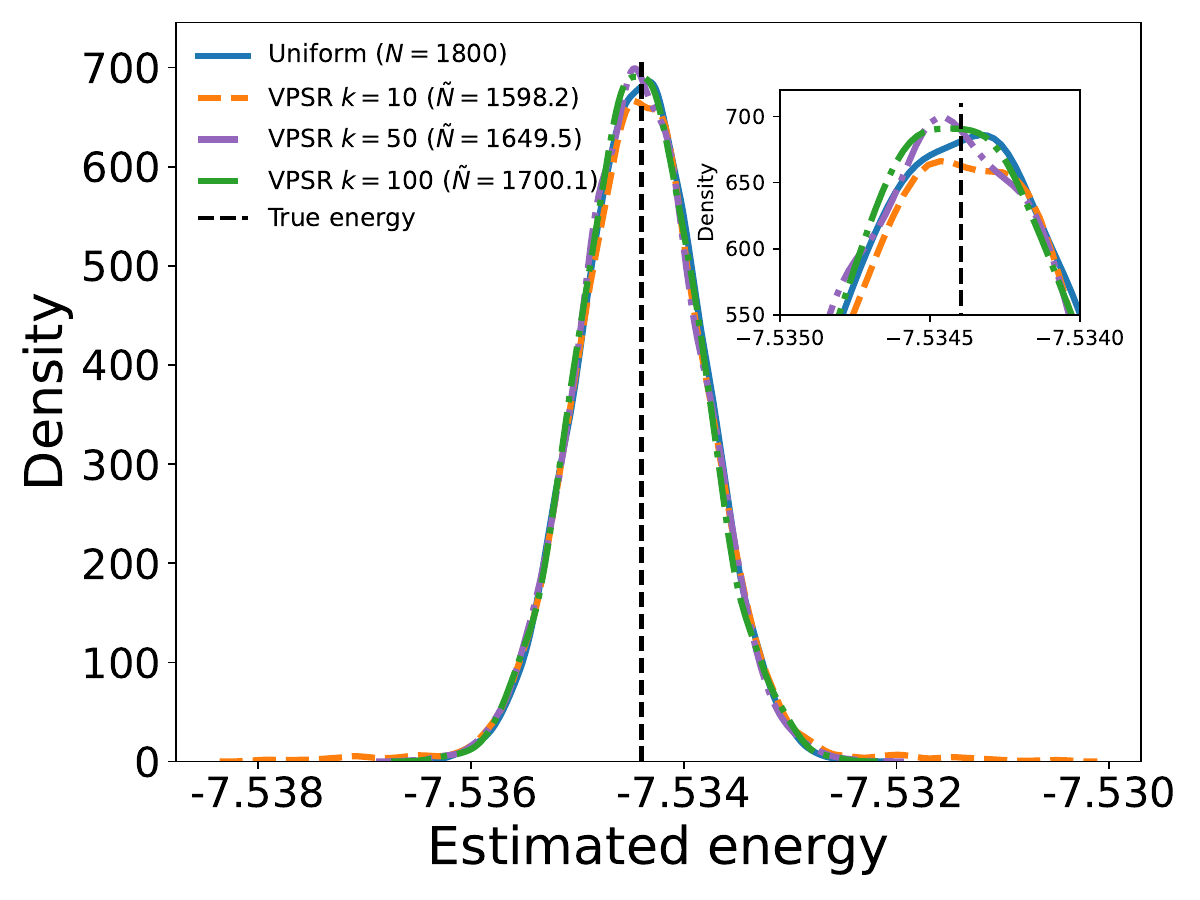}}
{\includegraphics[width=.45\textwidth]{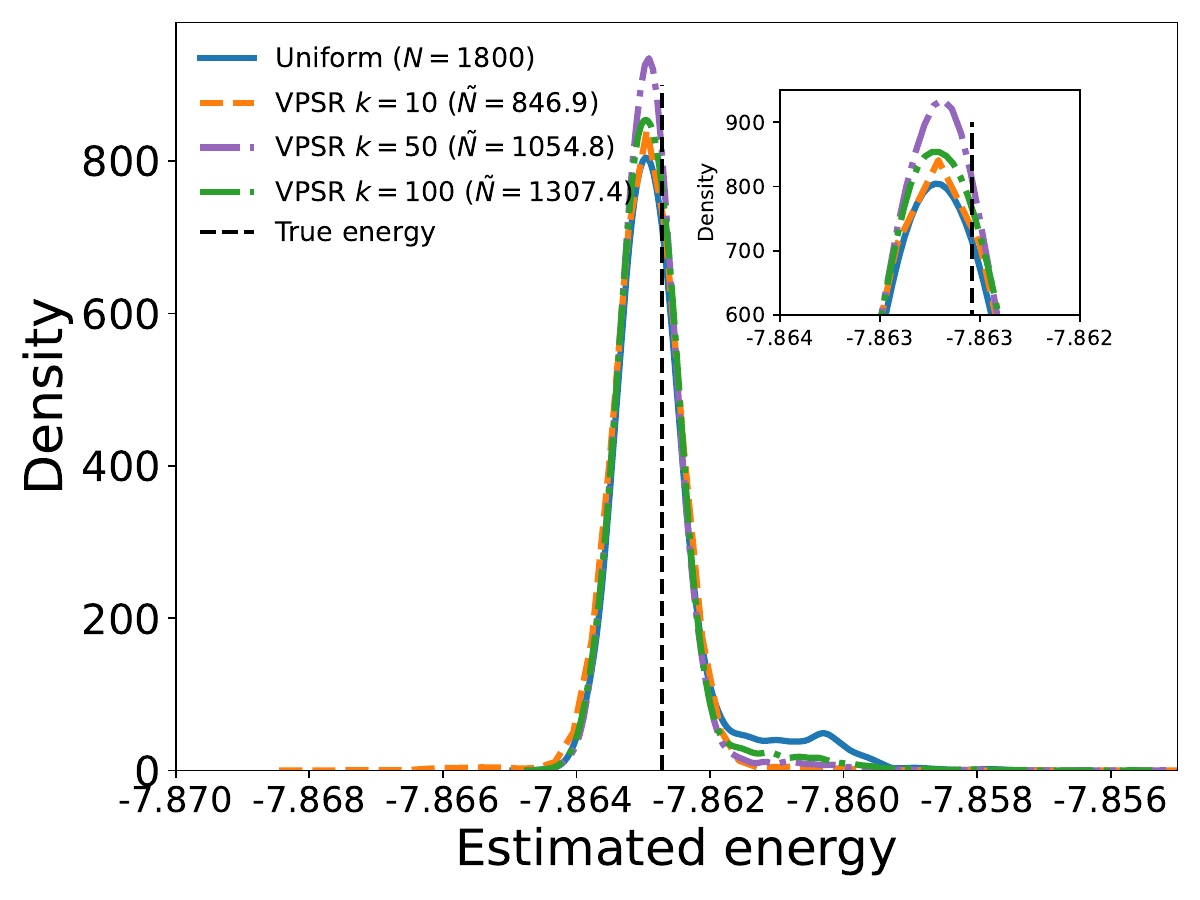}}
\caption{Distribution of the estimated energy expectation values of a LiH molecule for wavefunctions {prepared} with (left) $\vec\theta=\{0,0,0,0,0,0,0,0\}$ and (right) $\vec\theta=\{-0.032, 0.546, -1.587,  1.567,  0.015, -0.534, -1.570, -1.556\}$ based on 10,000 independent experiments.}\label{fig:energyestimationLiH}
\end{figure}

Figure~\ref{fig:energyestimationLiH} illustrates the impact of the sample size $k$ on both variance estimation and shot reduction by displaying the distributions of energy expectations measured from the quantum circuits defined by two $\vec \theta$ values for the LiH molecule. In these tests, we utilized a total shot budget of $N=1800$ and conducted 10,000 independent experiments. The two $\vec\theta$ values are: $\vec\theta=\{0,0,0,0,0,0,0,0\}$, representing the Hartree--Fock reference, and $\vec\theta=\{-0.032, 0.546, -1.587, 1.567, 0.015, -0.534, -1.570, -1.556\}$, representing the optimized ground state wavefunction.

Figure~\ref{fig:energyestimationLiH} demonstrates that of the three tested $k$ values, the smallest one, while offering the greatest reduction in shots, results in a noticeably broader distribution compared to the uniform assignment. Both $k=50$ and $k=100$ align closely with the uniform assignment for $\vec\theta=\{0,0,0,0,0,0,0,0\}$. However, at $\vec\theta=\{-0.032, 0.546, -1.587, 1.567, 0.015, -0.534, -1.570, -1.556\}$, $k=100$ provides a more accurate estimation of the energy expectation value's distribution. As a result, we will utilize $k=100$ for the VQE optimization process.

\begin{figure}[ht]
\centering\includegraphics[width=.99\textwidth]{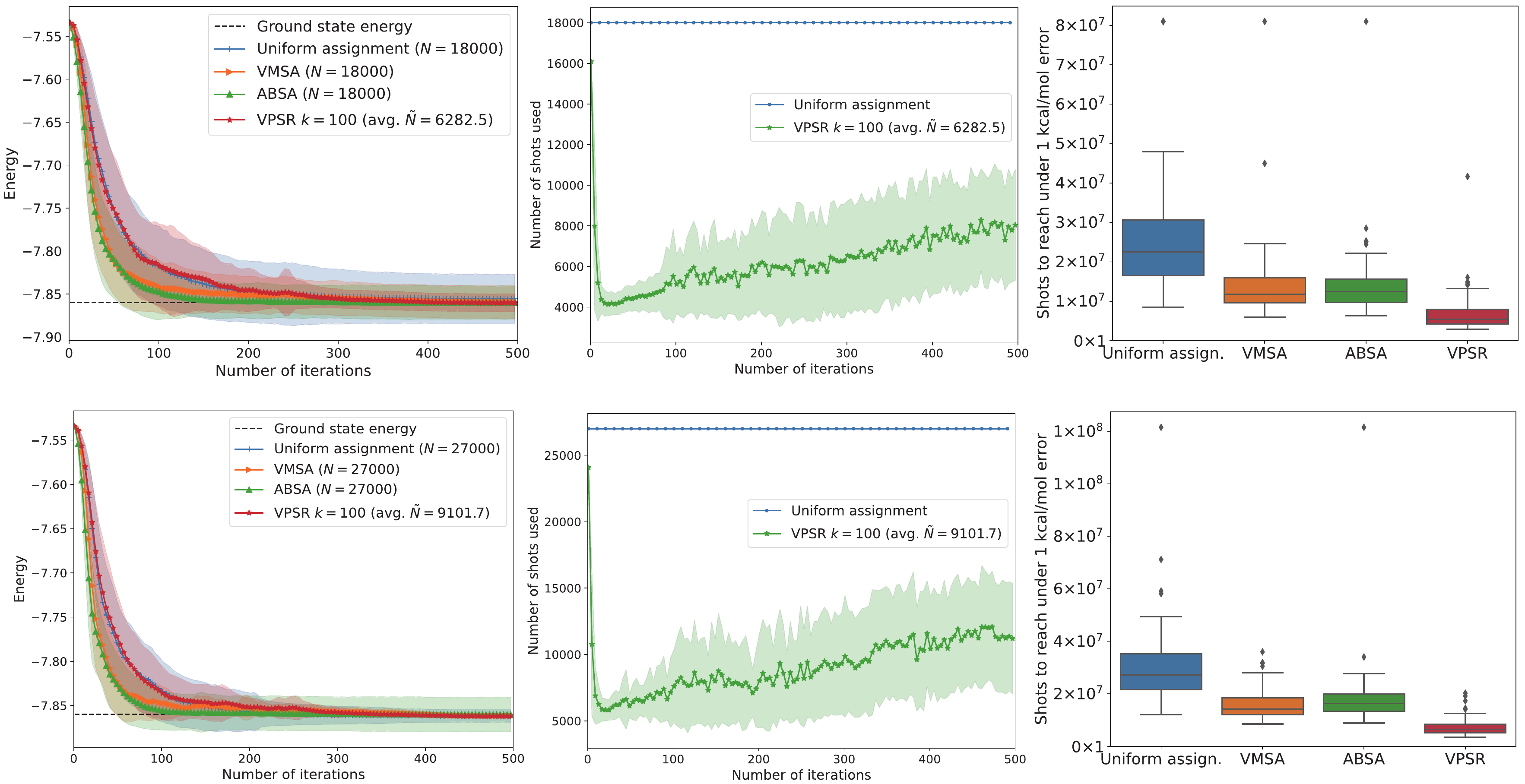}

\caption{Comparison of different shot assignment strategies for the VQE optimization of an eight-dimensional $\vec\theta$ parameter in the LiH ground state wavefunction. Top Panel: $N=18000$, $k=100$. Bottom Panel: $N=27000$, $k=100$. In the first column, the energy is plotted as a function of the number of iterations with the standard deviation indicated by a shaded area. The second column visualizes the distribution of shots across each iteration. The third column contains boxplots that summarize the number of shots needed to achieve an energy approximation within an error under 1.0 kcal/mol (1.6 mHartree) of the ground state energy. These boxplots facilitate a comparison of the distributional differences between datasets. Within each boxplot, the five horizontal lines, from bottom to top, represent the minimum, 25th percentile, median, 75th percentile, and maximum values. Black dots are used to denote outlier data points. The total number of shots utilized upon achieving the desired convergence: $N=18,000$ -- 
Uniform ($2.8\times10^{7}$),
ABSA ($1.5\times10^{7}$),
VMSA ($1.4\times10^{7}$),
VPSR ($7.4\times10^{6}$); 
$N=27,000$ --
Uniform ($3.1\times10^{7}$),
ABSA ($1.9\times10^{7}$),
VMSA  ($1.6\times10^{7}$),
VPSR  ($7.6\times10^{6}$).}
\label{fig:VQELiH}
\end{figure}

\begin{figure}[ht!]
\centering
{\includegraphics[width=.49\textwidth]{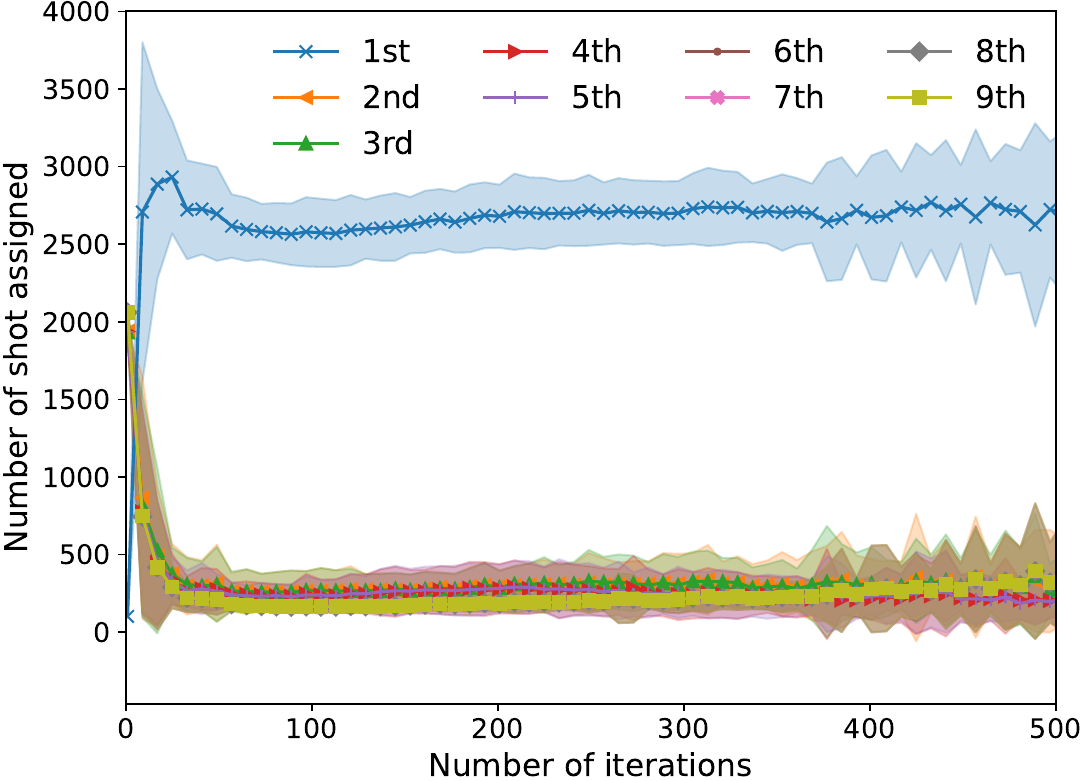}}
{\includegraphics[width=.49\textwidth]{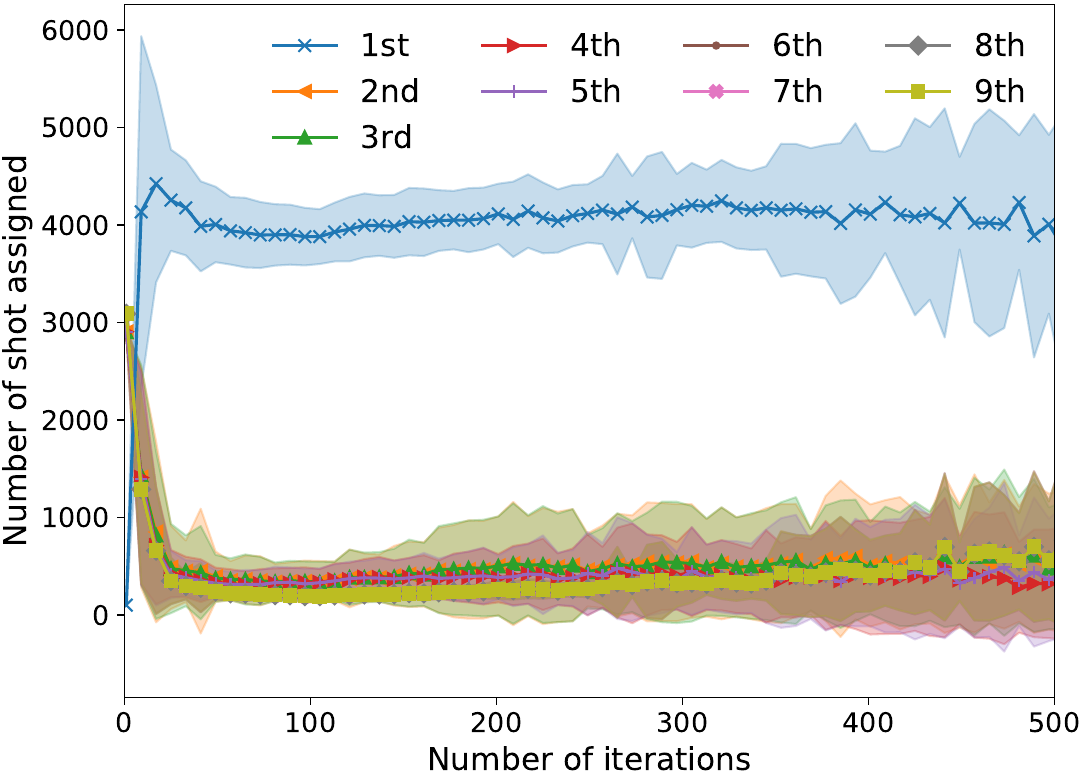}}
\caption{The number of shot assigned to each clique in VPSR for the VQE optimization of an eight-dimensional $\vec\theta$ parameter in the LiH ground state wavefunction. Left: $N=18000$, Right: $N=27000$. }\label{fig:LiHshotassignment}
\end{figure}

Figure \ref{fig:VQELiH} shows how the convergence of VQE optimization is affected by different shot assignment strategies, which allows us to  compare the effectiveness of the VPSR strategy with the VMSA and the non-variance-based (uniform and ABSA) approaches. The wavefunction initialization used $\vec\theta=(0, 0, 0, 0, 0, 0, 0, 0)$, corresponding to the Hartree--Fock reference. We explored two shot budgets: $N=18,000$ and $N=27,000$, with 60 independent experiments conducted for each. 

In this extensive eight-dimensional VQE optimization scenario, the VPSR approach yields a significant reduction in the number of measurement shots required to achieve chemical accuracy in the VQE optimization process -- 50\%--75\% fewer than the other three strategies tested. As illustrated in the first column of Figure \ref{fig:VQELiH}, the convergence of VQE optimization associaed with a VPSR shot assignment scheme  may not be as rapid as that of the VMSA or the ABSA approach in the initial 200 iterations. However, the overall shot reduction achieved upon VQE convergence is notably substantial in the VPSR based scheme, as indicated in the last column of Figure \ref{fig:VQELiH}. 
Although the ABSA based measurement scheme appears to enable faster convergence at the beginning of the VQE optimization procedure, it exhibits a much larger variance than the variance-minimization approaches. As a result, many more shots are used in the ABSA based measurements than the VPSR based approach in order to reach convergence to chemical accuracy.

The middle column reveals that VPSR's most significant variation in shot usage occurs early in the optimization process, followed by a consistent upward trend with minor fluctuations. This pattern highlights the adaptability of the VPSR approach, which effectively reduces the number of necessary shots in response to varying variances during the VQE optimization process.}
Figure \ref{fig:LiHshotassignment} illustrates the distribution of measurement shots across various cliques, highlighting that the first clique utilizes the highest number of shots. Moreover, the pattern of allocation shows significant fluctuation in the initial phase of the optimization process, stabilizing with minor variations as it nears convergence.

\section{Conclusion and Perspectives}\label{sec:conclusion}
{ 
In this work, we have developed a shot reduction strategy based on variance estimation, termed Variance-Preserved Shot Reduction (VPSR), which is designed to optimally manage the measurement budget. We have also provided mathematical derivations to elucidate its theoretical basis. Additionally, we have proposed an on-the-fly approach for variance estimation, which facilitates adaptive adjustment in shot allocation.

Numerical experiments were conducted to optimize the minimal basis wavefunctions of an H$_2$ molecule with a one-dimensional variational parameter and the wavefunction of a LiH molecule with an eight-dimensional variational parameter set, using the Variational Quantum Eigensolver (VQE). In these experiments, commuting terms were grouped into cliques for simultaneous measurements. The findings highlight that a small sample size in variance estimation can yield inaccurate energy expectation value distributions. However, with a well-tuned hyperparameter for choosing the sample size for variance estimation, the VPSR method effectively maintains a comparable optimization profile while significantly reducing the measurement budget, outperforming both uniform assignment and variance minimization strategies. In the extensive LiH test case, VPSR achieves a remarkable 65\% to 67\% reduction in shot measurements. The analysis also indicates that the VPSR's adaptability in modulating the optimal number of measurement shots in response to fluctuating variances is key to ensuring a smooth VQE optimization process.}

The proposed strategy's flexible shot allocation for different cliques makes it highly adaptable to complex and larger-scale quantum systems. Enhancements using more sophisticated techniques, like greedy algorithms and reinforcement learning,\cite{Li21_5482,liang2022finite,Li22_3169} could be employed to further optimize the assignment ratios. Future research endeavors could explore the effectiveness of these strategies in high-dimensional contexts, potentially broadening their applicability in various fields. 
By incorporating further analysis and mathematical validation, the VPSR method could be integrated with flexible grouping strategies, potentially enhancing its ability to reduce the number of shots required.

It is crucial to acknowledge that introducing adaptive variance estimation adds a level of complexity, affecting the parallelism of quantum measurements in each iteration. This impact is less severe compared to the VQE process, where a shift to classical hardware for optimization is required. While this reduction in parallelism might be offset by the decrease in measurement shots, its effectiveness in larger quantum simulations is yet to be confirmed. Given that measurement variance is heavily influenced by factors such as the quantum device type, its calibration, and inherent measurement noise, we recommend that users, in line with best practices for real-world quantum device applications, conduct preliminary variance testing to make an educated determination regarding the hyperparameter $k$.

\section*{Code Availability}
The source code for implementing the proposed VMSA and VPSR in VQE to estimate the ground state energy of the H$_2$ molecule is available in the online repository:  \url{https://github.com/LeungSamWai/OptimizingMeasurement}. The source code is released under the MIT license.

\section*{Acknowledgments:} 
Y.C. and X.L. acknowledge the support to develop reduced scaling computational methods from the Scientific Discovery through Advanced Computing (SciDAC) program sponsored by the Offices of Advanced Scientific Computing Research (ASCR) and Basic Energy Sciences (BES) of the U.S. Department of Energy (\#DE-SC0022263).
This material is based upon work supported by the U.S. Department of Energy, Office of Science, Office of Advanced Scientific Computing Research and Office of Basic Energy Science, Scientific Discovery through Advanced Computing (SciDAC) program under Contract No. DE-AC02-05CH11231 and the Accelerated Research for Quantum Computing Program under Contract No. DE-AC02-05CH11231.  The authors thank Dr. Shiv Upadhyay and Dr. Yuanran Zhu for helpful discussions on relevant topics.  This work used resources of the National Energy Research Scientific Computing Center (NERSC) using NERSC Award ASCR-ERCAP m1027 for 2023, which is supported by the Office of Science of the U.S. Department of Energy under Contract No. DE-AC02-05CH11231.

\pagebreak
\appendix

\section*{Appendix A: Cliques for H$_2$}
\label{appendix:H2}

In Ref.~\citet{o2016scalable} the utilization of the Bravyi-Kitaev(BK)-transformed UCC ansatz, along with spin symmetries, enables the mapping of the second-quantized Hamiltonian of the H$_2$ molecule onto 2 qubits with 6 terms.  This mapping gives rise to the following expression:
\begin{equation}
\hat{H} = g_{0} \mathbb{I} + g_{1} \mathbb{I} \otimes Z_0 + g_{2}Z_1 \otimes \mathbb{I} +g_{3}Z_{1} \otimes Z_{0} + g_{4}Y_{1}\otimes Y_{0} + g_{5}X_{1}\otimes X_{0},
\label{eq:H2_hamApp}
\end{equation}
where the amplitudes $\{g_i\}$ correspond to the integrals obtained from a Hartree-Fock calculation based on the bond distances and the type of atomic basis function. The relevant amplitudes $\{g_i\}$ are: 
{ $g_{0}=-0.5597$, $g_1=0.1615$, $g_2=-0.0166$, $g_3=0.4148$, $g_4=0.1226$, $g_5=0.1226$.~\cite{o2016scalable} }

We can consolidate the terms of H$_2$ into three cliques because the first three {non-identity terms ($g_{1-3}$)} commute with each other, enabling simultaneous measurement on both qubits.
\begin{equation}
  \begin{aligned}
   E_{1-3}(\theta) =& g_1 \braopket{\psi(\theta)}{ \mathbb{I} \otimes Z_0 }{\psi(\theta)} + g_2 \braopket{\psi(\theta)}{ Z_1 \otimes \mathbb{I} }{\psi(\theta)}\\ 
           &+g_3 \braopket{\psi(\theta)}{ Z_1 \otimes Z_0 }{\psi(\theta)}, \\     
  \end{aligned}    
\label{eq:E13_two_measure}
\end{equation}
where the same set of measurement results $|\braket{00}{\psi}|^2$, $|\braket{01}{\psi}|^2$, $|\braket{10}{\psi}|^2$, and $|\braket{11}{\psi}|^2$ can be used in:
\begin{equation}
  \begin{aligned}
   &g_1 \braopket{\psi(\theta)}{ \mathbb{I} \otimes Z_0 }{\psi(\theta)} \\
   =& g_1 \braopket{\psi}{ (\ket{0}\bra{0} -\ket{1}\bra{1})\otimes (\ket{0}\bra{0} +\ket{1}\bra{1}) }{\psi} \\
   =& g_1 (|\braket{00}{\psi}|^2 + |\braket{01}{\psi}|^2 - |\braket{10}{\psi}|^2 - |\braket{11}{\psi}|^2),
  \end{aligned}   
  \nonumber
\end{equation}
\begin{equation}
  \begin{aligned}
   &g_2 \braopket{\psi(\theta)}{ Z_1 \otimes \mathbb{I} }{\psi(\theta)} \\
   =& g_2 \braopket{\psi}{ (\ket{0}\bra{0} +\ket{1}\bra{1})\otimes (\ket{0}\bra{0} -\ket{1}\bra{1}) }{\psi} \\
   =& g_2 (|\braket{00}{\psi}|^2 - |\braket{01}{\psi}|^2 + |\braket{10}{\psi}|^2 - |\braket{11}{\psi}|^2),
  \end{aligned}   
  \nonumber
\end{equation}
and 
\begin{equation}
  \begin{aligned}
   &g_3 \braopket{\psi(\theta)}{ Z_1 \otimes Z_0 }{\psi(\theta)} \\
   =& g_3 \braopket{\psi}{ (\ket{0}\bra{0} -\ket{1}\bra{1})\otimes (\ket{0}\bra{0} -\ket{1}\bra{1}) }{\psi} \\
   =& g_3 (|\braket{00}{\psi}|^2 - |\braket{01}{\psi}|^2 - |\braket{10}{\psi}|^2 + |\braket{11}{\psi}|^2).
  \end{aligned}   
  \nonumber
\end{equation}
The other two cliques $E_4$ and $E_5$ are $g_4 \braopket{\psi(\theta)}{ Y_1 \otimes Y_0 }{\psi(\theta)}$ and $g_5 \braopket{\psi(\theta)}{ X_1 \otimes X_0 }{\psi(\theta)}$, respectively. 

While $E_4$ and $E_5$ in Eq. \ref{eq:H2_hamApp} require separate measurements, their measurement process can be streamlined through the use of appropriate transformation unitaries, which rotate them into the computational basis. Table~\ref{Tab:Transformation_two} outlines the unitaries needed for the {multi-qubits} measurement transformation for each Hamiltonian term. In this context, $H$ represents the Hadamard gate, and $S$ denotes an intrinsic quantum operation, which is equivalent to the square root of the Pauli $Z$ gate. { Note that the unitary transformation to the computational basis is typically not unique, and various transformation pathways may lead to different levels and types of measurement noise.} For $E_4$, as $({H}S^{\dagger})^{\dagger}Z({H}S^{\dagger})=Y$, we can apply ${H}S^{\dagger}$ to each qubit before $\ket{\psi(\theta)}$ and then measure it in the computational basis. Similarly, for $E_5$, as ${H}Z{H} = X$, we can apply ${H}$ to each qubit before measuring it.

\begin{table*}[htb]
\caption{Unitary transformation for {multi-qubits} measurement.}
\centering

\begin{tabular}{p{0.196\linewidth}p{0.78\linewidth}}
\hline
$\hat{H}_i$ & Transformation Unitaries\\
\hline
$ \mathbb{I} \otimes Z_0$ & ~~~~~~~~~~~~~~~~~~~~~~~~~NA \\
$  Z_1 \otimes \mathbb{I}$ & ~~~~~~~~~~~~~~~~~~~~~~~~~NA \\
$  Z_1 \otimes Z_0$ &  ~~~~~~~~~~~~~~~~~~~~~~~~~NA \\
$  Y_1 \otimes Y_0$ & $(H_1S_1^{\dagger} \otimes H_0S_0^{\dagger})^{\dagger}
          \cdot (Z_1 \otimes Z_0)\cdot  (H_1S_1^{\dagger} \otimes H_0S_0^{\dagger})$\\
$  X_1 \otimes X_0$ & $(H_1 \otimes H_0)^{\dagger}
          \cdot (Z_1 \otimes Z_0)\cdot  (H_1 \otimes H_0)$\\
$  \mathbb{I} \otimes Y_2 \otimes \mathbb{I} \otimes Y_0$ & $  (\mathbb{I} \otimes H_2S_2^{\dagger} \otimes \mathbb{I} \otimes H_0S_0^{\dagger})^{\dagger}\cdot (\mathbb{I} \otimes Z_2 \otimes \mathbb{I} \otimes Z_0)\cdot(\mathbb{I} \otimes H_2S_2^{\dagger} \otimes \mathbb{I} \otimes H_0S_0^{\dagger})$\\
$  \mathbb{I} \otimes X_2 \otimes \mathbb{I} \otimes X_0$ & $  (\mathbb{I} \otimes H_2 \otimes \mathbb{I} \otimes H_0)^{\dagger}\cdot (\mathbb{I} \otimes Z_2 \otimes \mathbb{I} \otimes Z_0)\cdot(\mathbb{I} \otimes H_2 \otimes \mathbb{I} \otimes H_0)$\\
$ Y_3  \otimes \mathbb{I} \otimes Y_1 \otimes \mathbb{I}$ & $  (H_3S_3^{\dagger} \otimes \mathbb{I} \otimes H_1S_1^{\dagger} \otimes \mathbb{I})^{\dagger}\cdot (Z_3 \otimes \mathbb{I} \otimes Z_1 \otimes \mathbb{I})\cdot(H_3S_3^{\dagger} \otimes \mathbb{I} \otimes H_1S_1^{\dagger} \otimes \mathbb{I})$\\
$ X_3  \otimes \mathbb{I} \otimes X_1 \otimes \mathbb{I}$ & $  (H_3 \otimes \mathbb{I} \otimes H_1 \otimes \mathbb{I})^{\dagger}\cdot (Z_3 \otimes \mathbb{I} \otimes Z_1 \otimes \mathbb{I})\cdot(H_3 \otimes \mathbb{I} \otimes H_1 \otimes \mathbb{I})$\\
$ X_3  \otimes X_2 \otimes Y_1 \otimes Y_0$ & $  (H_3 \otimes H_2 \otimes H_1S_1^{\dagger} \otimes H_0S_0^{\dagger})^{\dagger}\cdot (Z_3 \otimes Z_2 \otimes Z_1 \otimes Z_0)\cdot(H_3 \otimes H_2 \otimes H_1S_1^{\dagger} \otimes H_0S_0^{\dagger})$\\
$ Y_3  \otimes Y_2 \otimes X_1 \otimes X_0$ & $  (H_3S_3^{\dagger} \otimes H_2S_2^{\dagger} \otimes H_1 \otimes H_0)^{\dagger}\cdot (Z_3 \otimes Z_2 \otimes Z_1 \otimes Z_0)\cdot(H_3S_3^{\dagger} \otimes H_2S_2^{\dagger} \otimes H_1 \otimes H_0)$\\
$ Y_3  \otimes X_2 \otimes X_1 \otimes Y_0$ & $  (H_3S_3^{\dagger} \otimes H_2 \otimes H_1 \otimes H_0S_0^{\dagger})^{\dagger}\cdot (Z_3 \otimes Z_2 \otimes Z_1 \otimes Z_0)\cdot(H_3S_3^{\dagger} \otimes H_2 \otimes H_1 \otimes H_0S_0^{\dagger})$\\
$ X_3  \otimes Y_2 \otimes Y_1 \otimes X_0$ & $  (H_3 \otimes H_2S_2^{\dagger} \otimes H_1S_1^{\dagger} \otimes H_0)^{\dagger}\cdot (Z_3 \otimes Z_2 \otimes Z_1 \otimes Z_0)\cdot(H_3 \otimes H_2S_2^{\dagger} \otimes H_1S_1^{\dagger} \otimes H_0)$\\
\hline
\label{Tab:Transformation_two}
\end{tabular}
\end{table*}

\section*{Appendix B: Optimal Solution of the Objective Eq. \ref{eqn:objective2}}
\label{appendix:objective2}

A total of $N$ shots are allocated across $m$ terms, satisfying $\sum_{i=1}^mN_i=N$, where $N_i$ represents the number of shots assigned to the $i$-th term. The shots are distributed based on the proportion $\gamma_i>0$ and $N_i=N\gamma_i$, ensuring that $\sum_{i=1}^m \gamma_i=1$. Given the convergence threshold $\delta$, we have the following relationship with the empirical variance $\sigma_i^2(\vec\theta)$:
\begin{align*}
\delta\geq \sum_{i=1}^m\frac{\sigma_i(\vec\theta)^2}{N_i}=\frac{1}{N}\sum_{i=1}^m\frac{\sigma_i(\vec\theta)^2}{\gamma_i}&=\frac{1}{N}\sum_{i=1}^m\frac{\sigma_i(\vec\theta)^2}{\gamma_i}\sum_{i=1}^m \gamma_i \quad \text{(since $\sum_{i=1}^m \gamma_i=1$)}
\\&\geq \frac{1}{N} \left(\sum_{i=1}^m \sqrt{\frac{\sigma_i(\vec\theta)^2}{\gamma_i}}\sqrt{\gamma_i}\right)^2 \quad  \text{(Cauchy-Schwarz inequality)}
\\&=\frac{1}{N} \left(\sum_{i=1}^m \sigma_i(\vec\theta)\right)^2 
\end{align*}
Hence, we have the inequality 
\begin{equation}
N\geq \frac{\left(\sum_{i=1}^m \sigma_i(\vec\theta)\right)^2}{\delta}.\notag
\end{equation}
Equality holds if and only if the following condition is met: 
\begin{equation}
    \sqrt{\frac{\sigma_1(\vec\theta)^2}{\gamma_1}}/\sqrt{\gamma_1}=\cdots=\sqrt{\frac{\sigma_m(\vec\theta)^2}{\gamma_m}}/\sqrt{\gamma_m}\notag
\end{equation}
which can be expressed as 
\begin{equation}
\sigma_1(\vec\theta):\cdots:\sigma_m(\vec\theta)=\gamma_1(\vec\theta):\cdots:\gamma_m(\vec\theta)\notag 
\end{equation}

\section*{Appendix C: Cliques for LiH}
\label{appendix:LiH-cliques}
{Qubit-wise commutativity enables the simultaneous measurement of Pauli strings by ensuring that, for each qubit, the corresponding Pauli operators are either identical or including an identity operator.~\cite{yen2020measuring} Based on the principle, the Pauli strings in Eq.~\ref{eqn:lih_hamiltonian} can be grouped into 9 cliques:
\begin{itemize}
    \item $\{\mathbb{I} \otimes \mathbb{I} \otimes \mathbb{I} \otimes Z_0,~\mathbb{I} \otimes \mathbb{I} \otimes Z_1 \otimes Z_0,~\mathbb{I} \otimes \mathbb{I} \otimes Z_1 \otimes \mathbb{I},~\mathbb{I} \otimes Z_2 \otimes \mathbb{I} \otimes Z_0,~\mathbb{I} \otimes Z_2 \otimes Z_1 \otimes \mathbb{I},
    $\\{}
    $~\mathbb{I} \otimes Z_2 \otimes \mathbb{I} \otimes \mathbb{I},~Z_3 \otimes Z_2 \otimes \mathbb{I} \otimes \mathbb{I},~Z_3 \otimes \mathbb{I} \otimes Z_1 \otimes \mathbb{I},~Z_3 \otimes \mathbb{I} \otimes \mathbb{I} \otimes Z_0,~Z_3 \otimes \mathbb{I} \otimes \mathbb{I} \otimes \mathbb{I}\}$
    \item $\{\mathbb{I} \otimes Y_2 \otimes \mathbb{I} \otimes Y_0,~\mathbb{I} \otimes Y_2 \otimes Z_1 \otimes Y_0,~Z_3 \otimes Y_2 \otimes Z_1 \otimes Y_0\}$
    \item $\{\mathbb{I} \otimes X_2 \otimes \mathbb{I} \otimes X_0,~\mathbb{I} \otimes X_2 \otimes Z_1 \otimes X_0,~Z_3 \otimes X_2 \otimes Z_1 \otimes X_0\}$
    \item $\{Y_3 \otimes \mathbb{I} \otimes Y_1 \otimes \mathbb{I},~Y_3 \otimes Z_2 \otimes Y_1 \otimes Z_0,~Y_3 \otimes Z_2 \otimes Y_1 \otimes \mathbb{I}\}$
    \item $\{X_3 \otimes \mathbb{I} \otimes X_1 \otimes \mathbb{I},~X_3 \otimes Z_2 \otimes X_1 \otimes Z_0,~X_3 \otimes Z_2 \otimes X_1 \otimes \mathbb{I}\}$
    \item $\{X_3 \otimes X_2 \otimes Y_1 \otimes Y_0\}$
    \item $\{Y_3 \otimes Y_2 \otimes X_1 \otimes X_0\}$
    \item $\{Y_3 \otimes X_2 \otimes X_1 \otimes Y_0\}$
    \item $\{X_3 \otimes Y_2 \otimes Y_1 \otimes X_0\}$
\end{itemize}
Coefficients used for the LiH Hamiltonian at $R = 1.45$ \AA~are:

\begin{center}
\begin{tabular}{l|l|l}
\hline
Coefficient & Coefficient & Coefficient \\
\hline
\( g_0 = -7.4989469 \) & \( g_1 = -0.0029329 \) & \( g_2 = 0.0029329 \) \\
\hline
\( g_3 = 0.0129108 \) & \( g_4 = -0.0013743 \) & \( g_5 = 0.0115364 \) \\
\hline
\( g_6 = 0.0029329 \) & \( g_7 = -0.0029320 \) & \( g_8 = 0.0129108 \) \\
\hline
\( g_9 = -0.0013743 \) & \( g_{10} = 0.0115364 \) & \( g_{11} = 0.1619948 \) \\
\hline
\( g_{12} = 0.0115364 \) & \( g_{13} = 0.0115364 \) & \( g_{14} = 0.1244477 \) \\
\hline
\( g_{15} = 0.0541304 \) & \( g_{16} = 0.0570634 \) & \( g_{17} = 0.0129108 \) \\
\hline
\( g_{18} = -0.0013743 \) & \( g_{19} = 0.0129107 \) & \( g_{20} = -0.0013743 \) \\
\hline
\( g_{21} = 0.1619948 \) & \( g_{22} = 0.0570634 \) & \( g_{23} = 0.0541304 \) \\
\hline
\( g_{24} = -0.0132437 \) & \( g_{25} = 0.0847961 \) & \( g_{26} = -0.0132436 \) \\
\hline
\end{tabular}
\end{center}}

\clearpage
\pagebreak
\bibliography{bibliography}
\end{document}